\documentclass[journal=jctcce,manuscript=article]{achemso}

\usepackage[version=3]{mhchem} 
\usepackage{amsmath,amssymb,amsthm, amsfonts, mathtools, dsfont} 
\usepackage{bbm,bm,tensor, braket}
\usepackage{eqnarray,array,enumerate}
\usepackage{csquotes}
\usepackage{siunitx}
\usepackage{booktabs}
\usepackage{graphicx,wrapfig, caption, float, subcaption, epstopdf, setspace}
\usepackage{hyperref}
\usepackage{longtable}
\usepackage[section]{placeins}
\usepackage{multicol, multirow}
\usepackage{tikz,tkz-euclide}
\usetikzlibrary{shapes.geometric,arrows,arrows.meta, calc, positioning, automata, shadows, backgrounds,decorations.markings,decorations.pathreplacing}
\usepackage{tkz-fct}
\usetikzlibrary{intersections}
\usepackage{pgfplots}
\pgfplotsset{compat=1.9}
\usepgfplotslibrary{fillbetween}
\usepackage{pagecolor}\usepackage{amssymb}
\usepackage{algorithm, algpseudocode}
\usepackage{pdfpages}

\usepackage{cleveref}

\newcommand*{\citen}[1]{%
  \begingroup
    \romannumeral-`\x 
    \setcitestyle{numbers}%
    \cite{#1}%
  \endgroup   
}

\algrenewcommand\algorithmiccomment[1]{\hfill #1}

\def\br{\ensuremath\bm{r}}
\def\emp2{\ensuremath{E_{\text{corr}}^{\text{MP}2}}}

\author{Arno F{\"o}rster}
\email{a.t.l.foerster@vu.nl}
\affiliation{Theoretical Chemistry, Vrije Universiteit, De Boelelaan 1083, NL-1081 HV, Amsterdam, The Netherlands}
\author{Lucas Visscher}
\affiliation{Theoretical Chemistry, Vrije Universiteit, De Boelelaan 1083, NL-1081 HV, Amsterdam, The Netherlands}

\title{GW100: A Slater Type Orbital Perspective}


\DeclareUnicodeCharacter{2212}{-}
\begin{document}

\begin{abstract}
We calculate complete basis set (CBS) limit extrapolated ionization potentials (IP) and electron affinities (EA) with Slater Type Basis sets for the molecules in the GW100 database. To this end, we present two new Slater Type orbital (STO) basis sets of triple- (TZ) and quadruple-$\zeta$ (QZ) quality whose polarization is adequate for correlated-electron methods and which contain extra diffuse functions to be able to correctly calculate electron affinities of molecules with a positive Lowest Unoccupied Molecular Orbital (LUMO). We demonstrate, that going from TZ to QZ quality consistently reduces the basis set error of our computed IPs and EAs and we conclude that a good estimate of these quantities at the CBS limit can be obtained by extrapolation. With MADs from 70 to 85 meV, our CBS limit extrapolated ionization potentials are in good agreement with results from FHI-AIMS, TURBOMOLE, VASP and WEST while they differ by more than 130 meV on average from nanoGW. With a MAD of 160 meV, our electron affinities are also in good agreement with the WEST code. Especially for systems with positive LUMOs, the agreement is excellent. With respect to other codes, the STO type basis sets generally underestimate EAs of small molecules with strongly bound LUMOs. With 62 meV for IPs and 93 meV for EAs, we find much better agreement to CBS limit extrapolated results from FHI-AIMS for a set of 250 medium to large organic molecules. 
\end{abstract}

\section{\label{sec:introduction}Introduction}\protect

Over the last years, the $GW$ approximation (GWA)\cite{Hedin1965} has increasingly been applied to calculate charged excitations in finite systems.\cite{Blase2011, 
Faber2011,
Ren2012,
Bruneval2013,
VanSetten2013,
VanSetten2015,
Knight2016,
Golze2019} While formally more rigorous, fully self-consistent $GW$ (sc$GW$) calculations are relatively expensive and not necessarily very accurate for the calculation of ionization potentials (IP) and electron affinities (EA) for molecular systems.\cite{Caruso2016,Knight2016} Instead, the perturbative $G_0W_0$ approach is often the preferred alternative since it is computationally less demanding than sc$GW$ and can give significantly more accurate quasiparticle (QP) energies, provided that it is based on a  suitable reference.\cite{Bruneval2013, Caruso2016,Knight2016} 

Ideally, the result of a $G_0W_0$ calculation should be independent of the particular implementation of the method. In practice, however, choices regarding the numerical representation of the involved quantities must be made, including the choice of a single-particle basis as well as a discretization of frequency and/or time-variables. The choice of the single-particle basis also entail a choice regarding the representation of the core electrons as well as a treatment of virtual states. Both factors are decisive since it is known that core correlation plays a major role in $G_0W_0$ calculations\cite{VanSetten2015} but also since $GW$ QP energies converge very slowly to the complete basis set (CBS) limit.\cite{VanSetten2015, Golze2019, Stuke2020, Bruneval2020} For these reasons, achieving consensus between different $G_0W_0$ codes is challenging and requires careful convergence of a calculation with respect to all technical parameters. Due to limited resources and/or time constraints, it might not always be possible in applications to only work with converged parameters. In that case, one would like to know how a certain technical parameter affects the final result. 

For these reasons, comparison between different codes through systematic benchmarks is highly desirable. First, it allows to verify that the results from these codes agree within a reasonable margin of error. For $G_0W_0$ calculations, one usually aims for an accuracy of 100 meV. Second, such benchmarks are crucial in order to quantify the influence of the various technical parameters on the QP energies. Significant efforts in that direction have been initiated by van Setten et al.\cite{VanSetten2015} in 2015 with the publication of the GW100 database for finite systems. In their work, van Setten et al. compared the IPs and EAs of 100 small and medium-sized molecules on the $G_0W_0$@PBE level of theory, calculated with three different codes, the Gaussian type orbital (GTO) based all-electron code TURBOMOLE\cite{VanSetten2013, Balasubramani2020}, the numerical atomic orbital (NAO) based all-electron code FHI-AIMS\cite{FHIaims2009, Blum2009, Ren2012}, and the plane wave (PW) code BerkeleyGW.\cite{Hybertsen1986, Deslippe2012} Later, benchmarks for many more codes followed, including the PW implementations in VASP\cite{Kresse1993, Kresse1996, Kresse1996a, Liu2016} in 2017\cite{Maggio2017a} and WEST\cite{Govoni2015} in 2018,\cite{Govoni2018} and the real-space finite-element (RSFE) implementation in nanoGW\cite{Tiago2006} in 2019.\cite{Gao2019} Also the accuracy of many low-order scaling implementations of the $G_0W_0$ method were benchmarked against the GW100 database.\cite{Wilhelm2018, Wilhelm2021, Forster2020b, Duchemin2021} 

These studies established the choice of single-particle basis as a crucial factor causing major differences between different implementations. For instance, the results from TURBOMOLE, FHI-AIMS and MOLGW,\cite{Bruneval2016a} but also from the low-scaling implementations by Wilhelm et al.\cite{Wilhelm2021} in CP2K\cite{Kuhne2020a} and by Duchemin and Blase\cite{Duchemin2021}, all using the same def2-GTO type basis sets, agree within a few ten meV on average for GW100, even though these implementations differ in frequency treatment as well as calculation of four-center integrals. The differences between codes using different basis sets are considerably larger. The discrepancy between the TURBOMOLE and BerkeleyGW results of nearly 300 meV on average reported in ref.~\citen{VanSetten2015} for the Highest Occupied Molecular Orbital (HOMO) were not necessarily insightful since the BerkeleyGW results were not CBS limit extrapolated. With only around 60 meV on average, the agreement between the CBS limit extrapolated PW results obtained with VASP and TURBOMOLE was found to be significantly better.\cite{Maggio2017a} However, for EAs the disagreement between different codes is considerable larger and differences for systems with a positive LUMO can easily exceed several eV. It has also been pointed out in ref. ~\citen{Maggio2017a} that the type of GTO-type basis set has a major influence on these EAs and that Dunning's correlation consistent basis sets are more suitable than the def2-series which has been used in ref.~\citen{VanSetten2015}. Beside the choice of the basis set, the treatment of core electrons (pseudo-potentials vs. all-electron) also plays a decisive role for many systems.\cite{Govoni2018} 

Against this background, it seems that further benchmark results for GW100 using different basis set types are useful to advance the current understanding of the dependence of $GW$ QP energies on the basis set type. Recently, we presented the first production-level implementation of the $G_0W_0$ method with Slater type orbitals (STOs) in the Amsterdam density functional (ADF) module of the Amsterdam modelling suite (AMS).\cite{Snijders2001} The $G_0W_0$ implementation in ADF is detailed in ref.~\citen{Forster2020b} and is based on the space-time formulation of the $GW$ method first proposed by Godby and coworkers.\cite{H.N.Rojas1995, Rieger1999} It uses non-uniform imaginary time and imaginary frequency grids tailored to the system under investigation\cite{Kaltak2014, Kaltak2014a,Liu2016} and treats 4-point correlation functions with the pair-atomic density fitting (PADF) approximation, resulting in a low-order implementation with a very low prefactor.\cite{Wilhelm2021} To demonstrate the correctness of our implementation, we already presented IPs and EAs for the GW100 database in ref.~\citen{Forster2020b}. However, our results did not allow a meaningful comparison to implementations with other basis set types. Since it is not always straightforward to obtain converged minimax grids,\cite{Kaltak2014,
Wilhelm2018, 
Forster2020b} we used rather small imaginary frequency and time grids which were often not converged. Most importantly, in ref.~\citen{Forster2020b} we used standard Slater type basis sets optimized for independent-electron methods\cite{vanLenthe2003} which do not allow a systematic extrapolation to the CBS limit for correlated methods.\cite{Halkier1998, Jensen2013}

To be able to obtain accurate, CBS limit extrapolated IPs and EAs using STOS, we report here improvements over our original implementation regarding both parameters. For once, we implemented improved imaginary time and frequency grids which allows a systematic convergence to the limit of an infinite number of grid points. Second, and most importantly, we designed two new Slater type basis sets for all elements of the periodic table which we call (aug)-TZ3P and (aug)-QZ6P. They contain extra diffuse functions to be able to obtain accurate EAs for systems with LUMO above the vacuum level and in the choice of polarization functions we follow the requirements of correlated-electron methods\cite{Dunning1989, Jensen2013} as closely as possible: ADF only supports basis functions with angular momenta up to $l=3$ which is a clear restriction since already for second row atoms, a consistent polarization on the quadruple-$\zeta$ (QZ) level requires angular momenta up to $l=4$ and higher angular momenta functions are necessary for heavier elements.\cite{Jensen2013} For simplicity, we will refer to these basis sets as correlation consistent. However, we emphasize that they are not correlation consistent in a strict sense but rather as correlation consistent as possible in our current implementation. 

Despite these restrictions, as we will demonstrate in the following by comparison to results from other codes, the new basis sets allow for a reliable extrapolation to the CBS limit. Consequently, we present here CBS limit extrapolated IPs and EAs for the GW100 database with STOs. These results are the focus of the current work and are meant to complement the previous studies on GW100 using GTOs, PWs, and RSFEs. Since $G_0W_0$@PBE generally do not give accurate QP energies,\cite{Knight2016, Caruso2016, Rezaei2021} we will only focus on numerical aspects and abstain from comparison to experimental or high-level quantum chemistry reference values.\cite{Krause2015, Lange2018} To complement our results, we also calculate IPs and EAs of 250 molecules from the GW5000 database.\cite{Stuke2020} This work is organized as follows: In section~\ref{sec::theory} we shortly outline the $G_0W_0$ implementation in ADF and describe our new basis sets. We then present our IPs and EAs for GW100 and point out similarities and differences to other codes in section \ref{sec::Results}. Finally, section \ref{sec::conclusion} summarizes and concludes this work. 

\section{\label{sec::theory}Theory}
We start this section by reviewing shortly the $G_0W_0$ approximation (in the following, we refer to this approximation as GWA for simplicity) and describe the main features of the implementation in ADF. For more details we refer to ref.~\citen{Forster2020b}. Here, we mostly focus on the two factors which are most decisive for this work: The newly designed basis sets and the treatment of imaginary frequency and imaginary time.

\subsection{The $GW$ Approximation}
The $GW$ approximation is an approximation to the self-energy $\Sigma$ which appears in the Dyson equation.\cite{Hirata2017} It is defined as the difference between a dynamical one-body Hamiltonian $H$ and a static reference Hamiltonian $h$\cite{martin2016} corresponding to a non-interacting Fermi system (in practice, $h$ will be the Hamiltonian of generalized\cite{Seidl1996} KS\cite{Kohn1965, Hohenberg1964, EngelEberhardandDreizler2013} problem),
\begin{equation}
    \Sigma(1,2) = H(1,2) - h(1,2) \;.
\end{equation}
$H$ is defined by $H(\omega)= \omega - G(\omega)^{-1}$, where $G$ is the Green's function of the interacting Fermi-system. $1 = (\bm{r}_1,\sigma_1,\omega_1)$ collects space-, spin- and frequency coordinates of a particle but, as we do not consider spin-orbit coupling in the current work, we will omit spin for simplicity in the following. If one approximates $\Sigma$ as diagonal and real and transforms to a molecular orbital representation, the Dyson equation becomes
\begin{equation}
\label{quasi-particle-equations}
\omega_n = \braket{n|h|n} +
\braket{n| \operatorname{Re}\left(\Sigma(\omega_n)\right)|n}
\;,
\end{equation}
where $n$ labels a single-particle state. In case $n$ refers to the HOMO (LUMO) level, $-\omega_n$ is equal to the IP (EA) of the system. In the GWA, 
\begin{equation}
    \Sigma(1,2) = iG_0(1,2) * W_0(1,2) - v_{xc}(1,2) \;,
\end{equation}
where $*$ denotes convolution, $v_{xc}$ is the exchange-correlation potential of generalized KS-DFT, $W_0$ is the dynamically screened Coulomb interaction which is related to the bare Coulomb interaction $V$ via a Dyson equation,
\begin{equation}
\label{DysonW}
W_0(1,2) = V(1,2) + V(1,3)P_0(3,4)W_0(4,2) \;,
\end{equation}
and 
\begin{equation}
G_0(\omega) = (\omega-h)^{-1}   
\end{equation}
is the one-body Green's function corresponding to the non-interacting reference system. Integration over repeated indices is implied. The kernel of this Dyson equation, the polarizability $P_0$, is in turn calculated from the non-interacting Green's function in the random phase approximation (RPA)\cite{martin2016},
\begin{equation}
    P_0(1,2) = -i G_0(1,2) * G_0(2,1) \;.
\end{equation}
In principle, the RPA is distinct from the GWA but when referring to the $GW$ method in a quantum chemistry context it is usually implied that the RPA is made for $P_0$.

\begin{figure}[ht]
    \centering
    \includegraphics[width=0.7\textwidth]{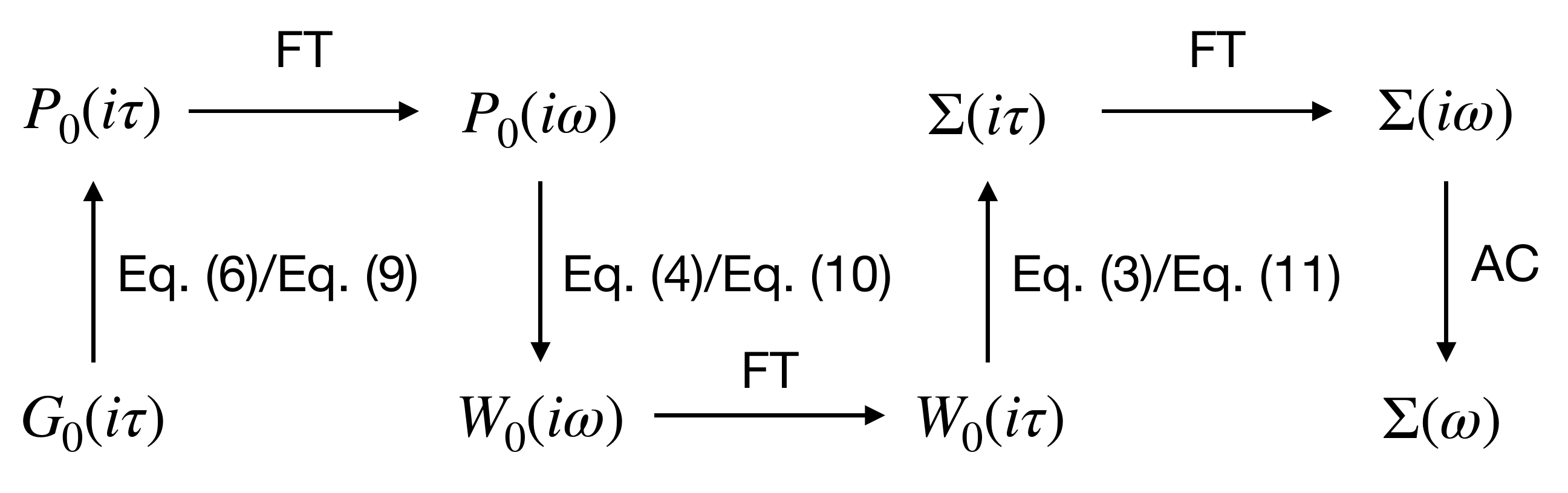}
    \caption{Schematic work flow for the calculation of the self-energy in the whole complex plane in ADF. FT denotes Fourier transform and AC denotes analytical continuation.}
    \label{fig::workflow}
\end{figure}

The $G_0W_0$ method in ADF is implemented in the space-time formalism introduced by Godby and coworkers.\cite{H.N.Rojas1995,Rieger1999} This means, eq. \eqref{DysonW} is solved in imaginary frequency, while $P$ and $\Sigma$ are calculated in imaginary time. From the imaginary time domain, $\Sigma$ is brought to the imaginary frequency axis. Since $\Sigma$ is analytic in the upper half plane, it can be analytically continued to the real axis. In practice this requires discrete grids to sample the imaginary time and imaginary frequency axes. In the absence of numerical noise we can use our knowledge of the self-energy on $N_{\omega}$ points on the imaginary frequency axis to interpolate the input data with a rational function using the Padé technique as outlined by Vidberg et al.\cite{{Vidberg1977}} The crucial assumption in this approach is that knowledge of $\Sigma$ at a small number of points on the imaginary frequency axis allows for an accurate interpolation. In case the single QP picure is valid, this assumption is reasonable for states in the vicinity to HOMO and LUMO energy since the interacting Greens-function has qualitatively the same structure as the non-interacting one (See for example ref.~\citen{Kopietz1997} or ref.~\citen{Mattuck1992}) and thus the self-energy (by definition) will be smooth.\cite{Cances2016} The $G_0W_0$ implementation in ADF follows the work flow shown in figure~\ref{fig::workflow}.

\subsection{Imaginary Time and Frequency Grids}

To discretize imaginary frequency and imaginary time in a suitable way, we follow Kresse and coworkers (building on earlier work by Almlöf\cite{Almlof1991} and Hackbusch and coworkers\cite{Takatsuka2008a}) and use non-uniform grids for both domains and switch between them using non-unifom Fourier transforms (FT).\cite{Kaltak2014,Kaltak2014a,Liu2016} The grid points are selected at run-time and are tailored to the system under investigation. Our implementation of these grids closely follows Kresse and coworkers\cite{Kaltak2014,Kaltak2014a,Liu2016} and is outlined in appendix~\ref{app::A}.

Similar grids in space-time method implementations have been used by other researchers as well\cite{Wilhelm2018, Wilhelm2021, Duchemin2021} to implement the $G_0W_0$ method for finite systems. Recently, Wilhelm et. al. benchmarked the convergence of QP energies in the GW100 database with respect to the grid sizes. In our older work,\cite{Forster2020b} our frequency grids were restricted to not more than 19 grid points and ref.~\citen{Wilhelm2021} clearly shows such grids to be insufficient to convergence all IPs and EAs in GW100. Consequently, we calculated new frequency grids which allow us to use up to 40 imaginary frequency points which ensures that the results are converged with respect to this parameter.\cite{Wilhelm2021} In ref.~\citen{Wilhelm2021}, Wilhelm et al. could also reproduce the results obtained by van Setten et al.\cite{VanSetten2015} with the TURBOMOLE code with an accuracy of a few meV. Since TURBOMOLE performs the frequency integration fully analytically, we conclude that the frequency integration is a numerical parameter which is well under control in our updated implementation. 

\subsection{Discretization of Real Space}
To discretize real space, we use a basis of $N_{AO}$ STOs, $\mathcal{S} = \left\{\chi_{\mu}\right\}_{\mu = 1, \dots \;, N_{AO}}$, which are related to the molecular orbitals (MO) $\phi$ by
\begin{equation}
    \label{expand}
    \phi_n = \sum_{\mu} b_{n\mu} \chi_{\mu}(\br) \;,
\end{equation}
All quantities which occur in \eqref{DysonW} are 4-point correlation functions and to this end, solution of \eqref{DysonW} would scale as $N^6$ when expanding these quantities in terms of $\mathcal{S}$. In the density-fitting (DF) approximation,\cite{Billingsley1971,
Baerends1973,
Whitten1973,
Sambe1975,
Dunlap1979,
Dunlap1979a,
Vahtras1993} one approximates the product basis of atomic orbitals $\mathcal{P} = \mathcal{S} \bigotimes \mathcal{S} $ with an auxiliary basis $\mathcal{A} = \left\{f_{\alpha}\right\}_{\alpha = 1, \dots \;, N_{aux}}$, where $N_{aux}$ grows linearly with system size, and introduces a basis transformation, 
\begin{equation}
    \label{dfeq}
    \mathcal{C}: \mathcal{P} \mapsto \mathcal{A} \;, \quad 
    \chi_{\mu}(\br)\chi_{\nu}(\br) = \sum_{p} c_{\mu \nu p}f_{p}(\br) \;.
\end{equation}
The equations which need to be solved now are
\begin{align}
\label{pFitting}
P_{pq,\tau} = & c_{\mu \nu p}P_{\mu \nu\kappa \lambda, \tau} c_{\kappa \lambda q} = 
-i c_{\mu \nu p} 
\left[G_0\right]_{\mu \kappa, \tau}
\left[G_0\right]_{\nu \lambda, \tau} c_{\kappa \lambda q} \\
\label{wFitting}
\left[W_0\right]_{pq,\omega} = & V_{pq} + V_{pr} \left[P_0\right]_{rs,\omega}
\left[W_0\right]_{sq,\omega}  \\
\label{scFitting}
\left[\Sigma_{xc}\right]_{ \mu \nu, \tau} =&
i\sum_{\kappa \lambda} \sum_{pq}
\left[G_0\right]_{\kappa \lambda, \tau}
c_{\mu \kappa q}\left[W_0\right]_{pq, \tau}
c_{\nu \lambda q} \;.
\end{align}
The computational bottlenecks in these equations are now the basis transformations in eqs. \eqref{pFitting} and \eqref{scFitting} which both scale as $N^4$ with system size. However, when the map $\mathcal{C}$ from eq. \eqref{dfeq} is constrained such that the number of non-zero-elements in $\mathcal{C}$ only grows linearly with system size, both equations only require $N^2$ operations.\cite{Wilhelm2018,Forster2020b} PADF is a way to achieve this goal by way of restricting the density fit to pairwise sums only and by subsequently introducing distance cut-offs. Also other schemes to introduce sparsity into $\mathcal{C}$ have been applied successfully to achieve low-order scaling $G_0W_0$ implementations\cite{Wilhelm2018,Wilhelm2021} (we refer to ref.~\citen{Wilhelm2021} for a detailed discussion of the pros and cons of different techniques), but PADF is arguably the best practical method to make the map $\mathcal{C}$ as sparse as possible.  Constructing $\mathcal{C}$ in that way only requires evaluation of a small number of integrals compared to other approaches\cite{Baerends1973} which makes this approach especially suitable for localized basis functions for which Coulomb integrals can not be calculated analytically, such as STOs or numerical atomic orbitals. In fact, the compute time spent on evaluating the remaining integrals in our implementation is negligible compared to the time needed to complete a $G_0W_0$ calculation. 

In PADF, only the fit functions centred on atoms $A$ and $B$ are used to expand a product of AOs which are centred on atoms $A$ and $B$. This translates into a matrix $\mathbf{c}$ for which $c_{\mu \nu p} \neq 0$ only if $\mu,\nu,p \in (A,B)$. One further introduces a cut-off so that for 2 AOs $\mu$, $\nu$ centred on atoms far away from each other, all corresponding matrix elements $c_{\mu \nu p}$ are set to zero. Employing PADF, standard auxiliary basis sets designed for global density fitting are not suitable (they are simply too small) and dedicated fit sets are needed. For an in-depth discussion of challenges in the design of these fit sets for correlated-electron methods we refer the reader to ref.~\citen{Forster2020b} and for a more explicit description of the fit sets, a list of exponents of the fit functions and more technical details to ref.~\citen{Forster2020}. 

\subsection{Slater Type Orbital Basis Sets}

In eq. \ref{expand}, $\mu$ is a  composite index, collecting the five defining parameters of a Slater type function
\begin{equation}
    \label{STO}
    \chi_{\mu}(r,\theta,\phi) = 
    \chi_{A,\alpha,n,l,m}(r,\theta,\phi) = 
    C(\alpha,n) r^{n-1} e^{-\alpha r} Z_{lm}(\theta,\phi) \;, \quad r = |\bm{r}-\bm{R}_A| \;,
\end{equation}
the exponent $\alpha$ and the quantum numbers $(n,l,m)$, as well as the nucleus $A$ on which the function is centred. In \eqref{STO}, $C(\alpha,n)$ is a normalization constant and $Z_{lm}$ denotes a real spherical harmonic. The main difference to Gaussian type functions is in the dependence on $r$ in the exponential instead of $r^2$ which mainly results in a different behavior close to the atomic nuclei and a slower decay for large $r$. While STO-type basis sets are well developed for independent-electron methods\cite{Cohen2002,vanLenthe2003, Watson2003, Watson2004,Chong2004} we are not aware of any attempt to construct general Slater type basis sets which are consistent with the requirements of correlated-electrons methods. Here, we make a first attempt to present such basis sets for the whole periodic table. The concept of correlation-consistency has also been applied to NAO basis sets for RPA total energy calculations.\cite{Zhang2013}

\subsubsection{Correlation Consistent Atomic Basis Sets}

In the construction of our basis sets we make two assumptions: First, we assume that the principles guiding the construction of GTO-type basis sets for correlated-electron methods which have been developed over the last decades should be applicable to STO-type basis sets as well. This can be justified as follows: Based on numerical experiments on correlation consistent GTO-type basis sets, Helgaker et al.\cite{Helgaker1997} established the relation
\begin{equation}
\label{helgakerExpression}
    E^{corr}_{\infty} - E^{corr}_{X} = aX^{-3}
\end{equation}
between the correlation energy $E^{corr}_{\infty}$ at the CBS limit and the correlation energy $E^{corr}_{X}$ calculated with a given correlation consistent basis set with cardinal number $X = l_{max} + 1$. Conceptually, their formula is based on earlier work by Schwartz\cite{Schwartz1962} and the mathematical more rigorous work by Hill\cite{Hill1985} on the convergence of the ground state of the Helium atom in a full configuration interaction calculation with respect to the single-particle basis. Later on, Kutzelnigg and Morgan\cite{Kutzelnigg1992} generalized that result to arbitrary $n$-electron systems for MP2 calculations. 

In principle, \eqref{helgakerExpression} is only valid in the limit of large $X$, however, there is numerical evidence that it is already a good approximation for $X=3$ and $X=4$.\cite{Helgaker1997,Halkier1998} Since \eqref{helgakerExpression} is independent of the type of localized basis functions\cite{Schwartz1962,Hill1985,Kutzelnigg1992}, this should also be the case for STO-type basis sets provided that they are also constructed in a correlation consistent fashion as first defined by Dunning;\cite{Dunning1989} such that the total basis set incompleteness error is distributed equally between the different angular momenta functions. This requires that polarization functions are added in well-defined sequences.\cite{Jensen2013} There is numerical evidence\cite{Jensen2013} that the consistent polarization for correlated methods is $1(l_{occ}+1)$ on the double-$\zeta$ (DZ), $2(l_{occ}+1)1(l_{occ}+2)$ on the TZ and $3(l_{occ}+1)2(l_{occ}+2)(l_{occ}+3)$ on the QZ level, where $l_{occ}$ denotes the angular momentum of the highest occupied orbital in an atom. This requirement is usually only followed strictly for the first three rows of the periodic table\cite{Jensen2013} and the design of correlation consistent basis sets for heavier elements might follow different principles.\cite{Dyall1998,Dyall2002,Dyall2004,Dyall2006,Gomes2010}

\subsubsection{Construction of Correlation Consistent Slater-Type Basis sets}

Based on the considerations above, we construct correlation consistent basis sets of TZ and QZ quality. We name these basis sets TZ3P and QZ6P, respectively. We also augment these basis sets with additional diffuse functions and we name these basis sets aug-TZ3P and aug-QZ6P, respectively. The acronym $xP$ refers to the number of polarization functions we use for the elements of the first three rows of the periodic table. This choice is consistent with the requirements for correlation consistent basis sets stated above. 

At this point, we introduce our second assumption: Since the ADF code only supports basis functions up to $l=3$, for second- and third-row elements we chose the polarization $2d1f$ and $3d3f$ for TZ and QZ respectively, and for consistency also $2p1d$ and $pd3d$ for Hydrogen and Helium. A good justification for the validity of this approximation can not be given and as we will see later, our results indeed suggest that the replacement of a $g$ with another $f$ functions negatively affects the QP energies for small molecules. Generalization of ADF to accommodate use of higher angular momenta functions is therefore desirable.

The TZ2P and QZ4P basis sets which have been described in detail in ref.~\cite{vanLenthe2003} serve as starting points for our new basis sets. TZ2P (QZ4P) is of DZ (TZ) quality in the core region and of TZ (QZ) quality in the valence region. For the TZ3P and QZ6P basis sets, we chose not to optimize exponents but rather to add additional polarization functions in an ad-hoc manner. This approach can be justified by the observation that precise values of exponents become less important as a basis set approaches completeness.\cite{Cohen2002,Jensen2013} This is especially true for molecules as opposed to isolated atoms. The latter case generally requires larger and more optimized basis sets than the molecular case. Instead, we rather focus on choosing the exponents in a way that their overlap is small and linear dependency problems are more likely to be avoided.

TZ3P is simply obtained by augmenting the TZ2P basis set by another $l_{occ}+1$-function for all elements. The exponent is chosen so that it is twice as large as the exponent of the $l_{occ}+1$-function in the TZ2P basis set. This is due to the fact that the exponents of the polarization functions in the TZ2P and QZ4P basis sets are chosen in a way that the basis sets become more accurate in the valence region, which is favorable for the calculation of bonding energies. The calculation of IPs also requires the accurate representation of the electron density closer to the core, and TZ3P should yield a major improvement over TZ2P in that respect. The same reasoning has also been followed in ref.~\citen{Autschbach2000}. In complete analogy, the QZ6P basis set is obtained by adding an additional tight $l_{occ}+1$-function and an $l_{occ}+2$-function for all elements. For Fluorine and Chlorine we added an additional shell of $d$-function since we found exceptionally slow convergence of QP energies to the CBS limit. This will be discussed in more detail in section~\ref{sec::Results}. 

The exponents $\alpha_{1},\alpha_{2}$ of the polarization functions in the QZ4P basis set fulfill $\frac{\alpha_{1}}{\alpha_{2}} \approx 2$ for each $l$. Thus, the polarization functions in the TZ3P and QZ6P basis sets loosely follow an even-tempering scheme\cite{Reeves1963a},
\begin{equation}
\label{evenTemper}
    \alpha_i = \alpha_1 \beta^{i-1}\;, \quad i = 2, \dots , M \;, \quad \beta = 2 \;,
\end{equation}
with M being equal to two (three) for $l_{occ}+1$ in TZ3P (QZ6P) and one (three) for $l_{occ}+2$. The value of $\beta$ is rather large to avoid linear dependency problems, and in conjunction with a rather small $\alpha_1$ ensures that the exponents span a rather wide range of values in the QZ6P case.\cite{Silver1978} Ren et al. recently used a similar reasoning to chose the exponents of Slater type functions for $G_0W_0$ calculations for periodic systems.\cite{Ren2021}

\subsubsection{Adding Diffuse Functions}

It is known from electron scattering experiments that molecular electron affinities are sometimes negative, i.e. their anion state at the geometry of the neutral molecule is unstable.\cite{Schulz1973, Jordan1978,Aflatooni1998} This corresponds to a positive LUMO QP energy which formally corresponds to a non-normalizable continuum orbital. However, as an artefact of working with a finite basis, the orbital will always be constrained to be normalizable.\cite{Tozer2005} Very diffuse functions are then needed to mimic the continuum state, and for this reason we augment our basis sets with additional diffuse functions (See also ref.~\citen{Kendall1992} and ref.~\citen{Zheng2011}). The resulting basis sets are denoted as aug-TZ3P and aug-QZ6P and are obtained from TZ3P and QZ6P, respectively by adding diffuse $s$, $p$ and $d$-functions for all element types, with the important exceptions of first row atoms where we only add diffuse $s$ and $p$ functions.\bibnote{This is slightly different from usual augmentation approaches encountered for Gaussian type basis sets. Dunning and coworkers defined the prefix "-aug" to mean adding an additional diffuse functions for all angular momenta already present in a basis set for every atom type.\cite{Kendall1992} On the contrary, in a process which they denoted as minimal augmentation, Truhlar and coworkers only added diffuse $s$ and $p$ functions to all elements heavier than Hydrogen.\cite{Zheng2011} Our approach is can be seen as a compromise between both.} We decided to chose the exponents of the diffuse functions in line with \eqref{evenTemper}, except for a small shift. More precisely, the exponent of the most diffuse function for a given angular momentum is $\alpha_{l,min}$, the exponents of the diffuse function is chosen to be
\begin{equation}
\label{diffuseEXP}
\alpha_{l,diffuse} = \alpha_{l,min}/2 + 0.05 \alpha_{l,min}   
\end{equation}
The exceptions are the elements Hydrogen to Beryllium for which we chose the already optimized exponents by Chong et al.\cite{Chong2005} These exponents are very close to fulfilling \eqref{diffuseEXP}. Our choice of exponents is a compromise between two requirements: We would like the additional functions to be as diffuse as possible but we still want to be able to fit them accurately with a large auxiliary basis set. As discussed for instance in ref.~\citen{Forster2020b}, this is already challenging. In practice, this means that $2\alpha_{l,diffuse}$ should be at least equal or preferably slightly larger (for this reason we added the shift in \eqref{diffuseEXP}) than the exponent of the most diffuse function in the auxiliary basis set. 

\section{\label{sec::Results}Results}
\subsection{Computational Details}
All calculations have been performed with a locally modified development version of ADF2020\cite{adf2020} using the implementation as described in ref.~\citen{Forster2020b} and using the updated imaginary frequency grids as described in this work.\bibnote{The code to generate the imaginary time grids can be found on Github. (\href{https://github.com/bhelmichparis/laplace-minimax}{https://github.com/bhelmichparis/laplace-minimax}). Please also see ref.~\citen{Takatsuka2008a} and \citen{Helmich-Paris2016}.}  

\subsubsection{GW100}
We follow the protocol for GW100 as described by van Setten et al.\cite{VanSetten2015} and perform non-relativistic $G_0W_0$@PBE\cite{Perdew1996,Perdew1996a} calculations. We use the structures as available on the webpage for the GW100 database and also use the updated structures for Phenol and Vynilbromide.\bibnote{Data downloaded from the webiste of the GW100 project by Van Setten et al., \href{https://gw100.wordpress.com}{https://gw100.wordpress.com}} For consistency with other benchmarks for GW100, we always use the QP energy obtained from the KS LUMO energy, which is usually, but not always, the energetically lowest virtual QP energy. For a detailed discussion of the effect of orbital reordering we refer the reader to ref.~\citen{Gao2019}. 

We performed calculations with the augmented versions of the basis sets described in this work and extrapolated them to the CBS limit as described for instance in ref.~\citen{VanSetten2013}: We calculate the QP energies $\epsilon_n$ with the aug-TZ3P and aug-QZ6P basis sets and estimate the CBS limit as
\begin{equation}
    \label{cbsExtra}
    \epsilon_n^{CBS} = \epsilon_n^{QZ} - 
    \frac{1}{N^{QZ}_{bas} }\frac{\epsilon_n^{QZ} - \epsilon_n^{TZ}}{\frac{1}{N^{QZ}_{bas}} - \frac{1}{N^{TZ}_{bas}}} \;,
\end{equation}
where $\epsilon_n^{QZ}$ ($\epsilon_n^{TZ}$) denotes the value of the QP energy using aug-QZ6P (aug-TZ3P) and $N^{QZ}_{bas}$ and $N^{TZ}_{bas}$ denote the respective numbers of basis functions. Since we work with spherical harmonics, there are 5 $d$ and 7 $f$ functions. In all calculations, we set the \texttt{numericalQuality} key to \texttt{Good} and used 32 imaginary time and 32 imaginary frequency points each. As explained in section~\ref{app::A}, this should be understood as the upper bound for the number of grid points. We report the number of grid points actually used in each calculation in the supporting information. Per default we also used the \texttt{Good} auxiliary basis set. However, for all systems in GW100 containing fourth- or fifth-row elements and for all systems smaller than 4 atoms we used the \texttt{Normal} auxiliary basis set to prevent issues with over-fitting (see ref.~\citen{Forster2020}). For systems with a positive LUMO we used the \texttt{veryGood} fit set. This fit set contains additional diffuse fit functions which are necessary to accurately represent the diffuse electron densities of these systems. These fit sets are discussed in ref.~\citen{Forster2020}. Also for Guanine, Uracil, Pentasilane and Ethoxyethane we used the \texttt{VeryGood} fit set since we observed inconsistencies with the \texttt{Good} fit set. I all calculations, we set \texttt{Dependency Bas=5e-4} in the AMS input as described in ref.~\citen{Forster2020b}.

\subsubsection{GW5000}
We also performed $G_0W_0$@PBE0\cite{Ernzerhof1999, Adamo1999} calculations for a subset of of 250 molecules from the GW5000 database\cite{Stuke2020} using the zeroth order regular approximation (ZORA).\cite{VanLenthe1993, VanLenthe1994, VanLenthe1996, VanLenthe1996a} Calculations are performed with the non-augmented TZ3P and QZ6P basis sets. Eq. \eqref{cbsExtra} is used for CBS limit extrapolation. We use 24 points in imaginary frequency and imaginary time each, \texttt{numericalQuality Good} and the \texttt{Normal} auxiliary basis set which is sufficient for non-augmented basis sets.\cite{Forster2020b} We set \texttt{Dependency Bas=5e-4}. 

\subsection{Basis Set Errors} 
Before we discuss in detail the comparison of the STO results to the ones from other codes, we consider the basis set errors and basis set convergence properties of aug-TZ3P and aug-QZ6P. Using eq. \eqref{cbsExtra}, one implicitly assumes that when going to a larger basis set, each additional basis function reduces the basis set error on average by the same amount. In other words, one assumes uniform convergence of the basis set expansion, a natural property of finite elements in real- or reciprocal space. It is clear that such an assumption will only be justified for rather large basis sets and, for the same reason, extrapolation is generally more reliable for larger systems. Usually one would also like to use three or even more data points instead of using \eqref{cbsExtra}. It has, however, been pointed out\cite{Halkier1998} that including a calculation with a basis set of quality lower than TZ will detoriate the quality of the fit. Therefore, we calculate QP energies at the CBS limit from TZ and QZ results only. 

\begin{figure}[ht]
    \centering
    \includegraphics[width=0.7\textwidth]{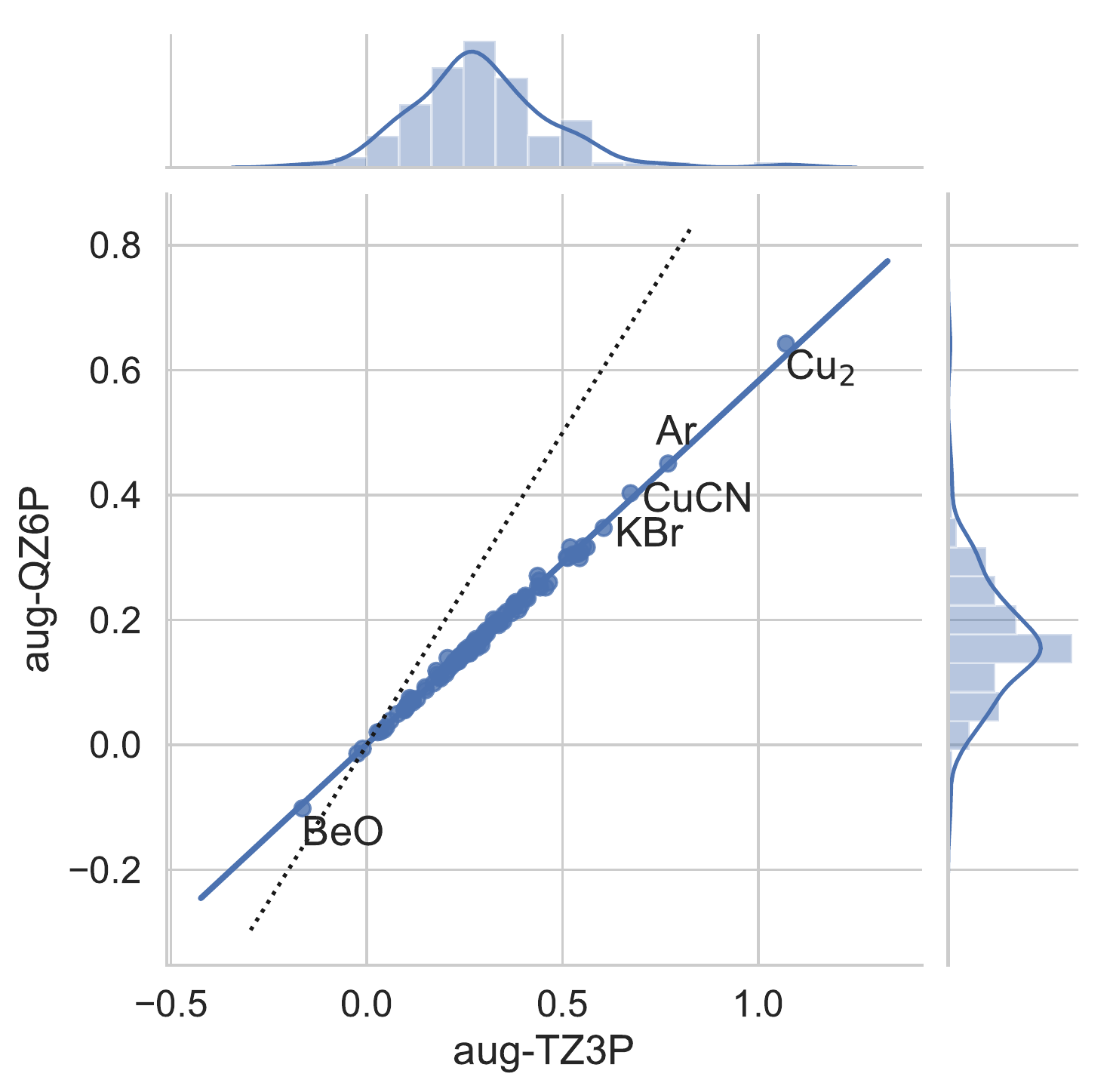}
\caption{Basis set errors with respect to the CBS limit extrapolated IPs in the GW100 database using the aug-TZ3P and aug-QZ6P basis sets. The univariate plots show the distributions of errors with respect to the CBS limit extrapolated values. All values are in eV.}
    \label{fig::bseHOMO}
\end{figure}

\begin{figure}[ht]
    \centering
    \includegraphics[width=0.7\textwidth]{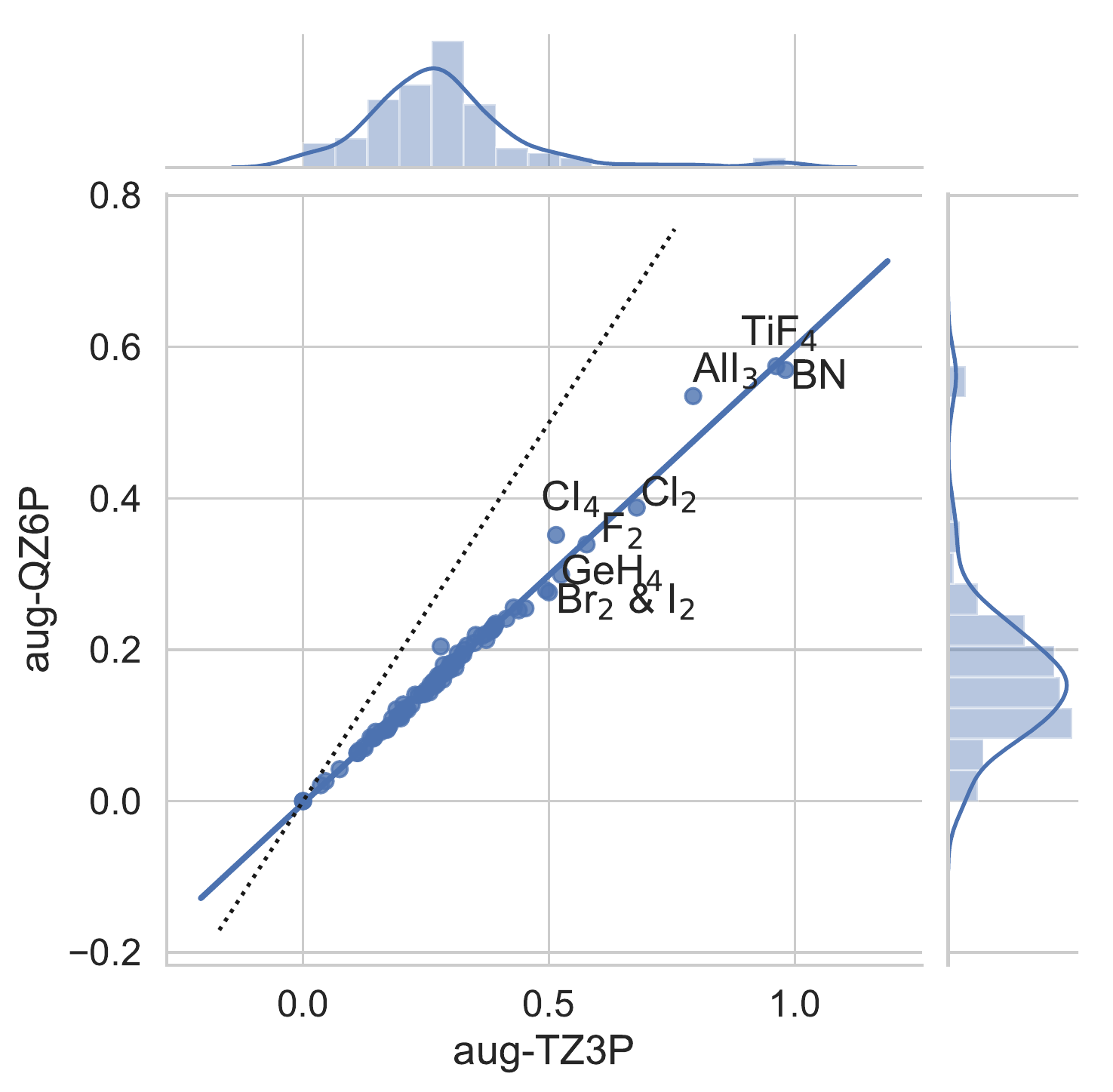}
\caption{Basis set errors with respect to the CBS limit extrapolated EAs in the GW100 database using the aug-TZ3P and aug-QZ6P basis sets. The univariate plots show the distributions of errors with respect to the CBS limit extrapolated values. All values are in eV.}
    \label{fig::bseLUMO}
\end{figure}

To demonstrate the accuracy of an extrapolation scheme, one would ideally compare the extrapolated result to one obtained with a very large basis which already gives a result very close to the CBS limit. Due to the limitations of our basis sets to angular momenta $\leq 3$, this is not possible for us. As a rule of thumb, extrapolation with basis sets of cardinality $X$ and $X-1$ can provide the accuracy of a calculation using a basis set of accuracy $X+1$.\cite{Jensen2013} Recently, Bruneval et al.\cite{Bruneval2020} found basis set errors of about 60 meV for the IPs of a large set of small to medium organic molecules with the cc-pV5Z basis set. This is of the same order as the typical accuracy in a photo-ionization experiment\cite{Bruneval2020} and considerably lower than the 150 meV for IPs which are usually found using the cc-pVQZ basis set.\cite{Stuke2020,Bruneval2020} For EAs, one can usually expect errors of the same order of magnitude than IPs when augmented basis sets are used.\cite{Wilhelm2016a} It is thus reasonable to use this number as an estimate of the average error in our extrapolation.

\begin{figure}[ht]
    \centering
    \includegraphics[width=0.7\textwidth]{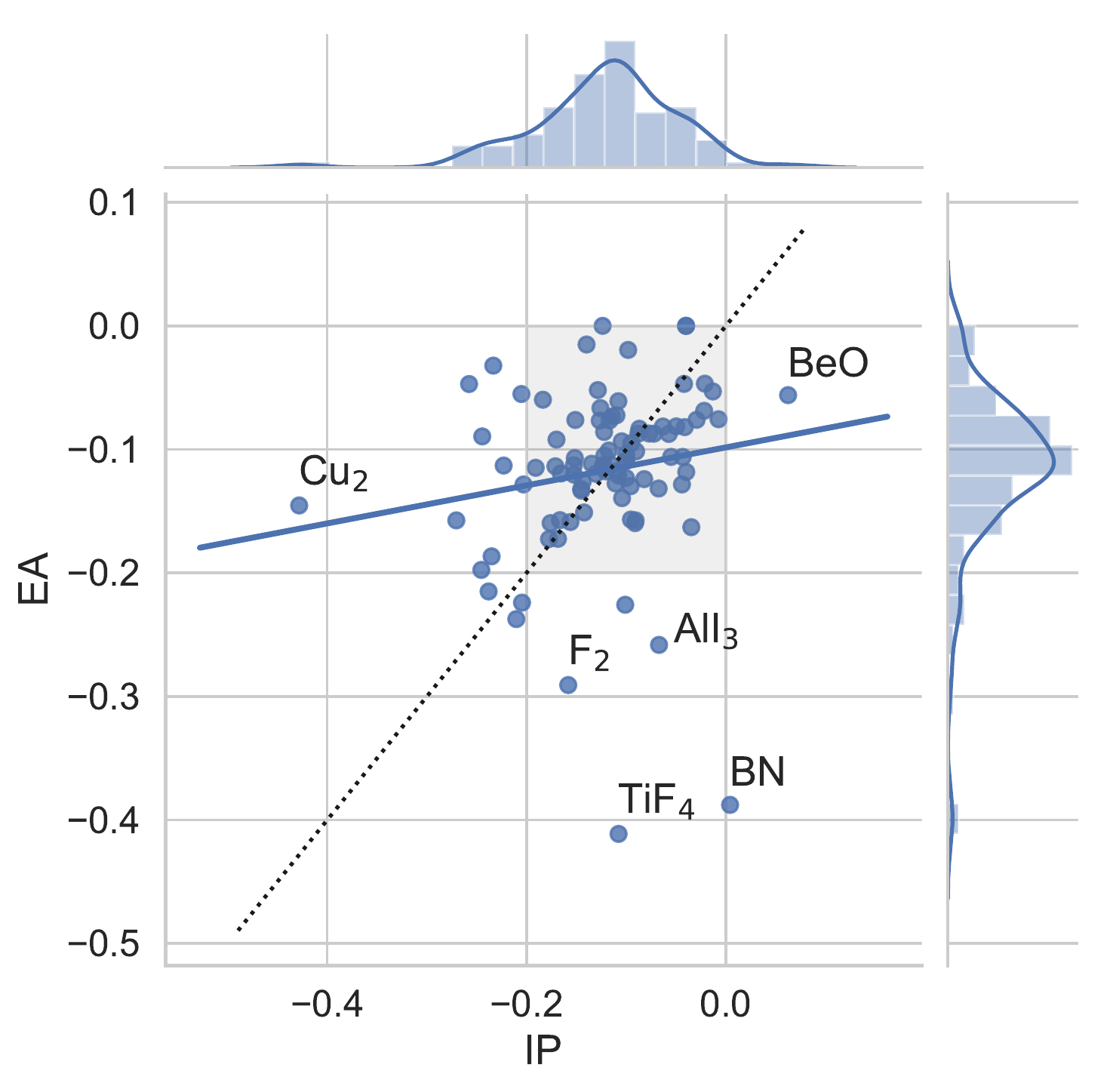}
    \caption{Correlation between differences between the aug-TZ3P and aug-QZ6P results $\Delta_{T-Q}$ for EAs (y-axis) and IPs (x-axis). The univariate plots show the corresponding distributions. The blue straight line is a linear fit and the dotted line is defined by $\Delta_{IP} = \Delta_{EA}$. All values are in eV.}
    \label{fig::IPcorrEA}
\end{figure}

The distributions of basis set errors with respect to the CBS limit extrapolated IPs and EAs (excluding noble gases and the hydrogen molecule in the latter) for the GW100 database are shown in figures~\ref{fig::bseHOMO} and~\ref{fig::bseLUMO}, respectively. The average basis set error reduces from 300 (290) meV to 170 (170) meV for IPs (EAs), i.e. there seems to be no qualitative difference in the convergence to the CBS limit for IP and EA as one would expect for augmented basis sets. We also notice that with one exception of the IP of \ce{BeO}, the basis set error on the TZ level is always lower than the one on the QZ level. In both plots, we also highlight some systems for which the convergence to the CBS limit seems to be rather slow, ie. The difference between QP energies on the TZ and QZ level are very large. For these systems, CBS limit extrapolation will be less accurate. Without exception, the problematic systems are composed of only a few atoms. 

Also a good correlation between the basis set errors for IPs and EAs is desirable since it implies a fast convergence of the HOMO-LUMO QP gap to the CBS limit. Fast convergence of this quantity with basis sets augmented with diffuse functions has been demonstrated before.\cite{Wilhelm2016a, Forster2020b} At first glance, the distributions of IP and EA BSEs appear rather similar, suggesting that such an error cancellation might also be found for our basis sets. To investigate this further, we plot all pairs $(\Delta(QZ-TZ)_{IP_i},\Delta(QZ-TZ)_{EA_i})$ (bivariate plot) together with the corresponding error distributions (univariate plots) in figure~\ref{fig::IPcorrEA} (omitting again all noble gases and \ce{H2}). The blue solid line is a linear fit, the dotted black line is defined by the equation $\Delta(QZ-TZ)_{IP_i} = \Delta(QZ-TZ)_{EA_i}$, and Gaussian kernels are fitted to the univariate distributions. A few molecules with large BSE for the IP but small basis set error for the EA (\ce{Cu2}) or vice-versa (\ce{BN}, \ce{TiF4}, \ce{F2}, \ce{AlI3}) aside, most systems cluster around the dotted line in the grey shaded area in which IP and EA BSEs should cancel each other to a large extent.

\subsection{Comparison to GTO-Type Basis Sets}

\begin{figure}[ht]
    \centering
    \includegraphics[width=0.7\textwidth]{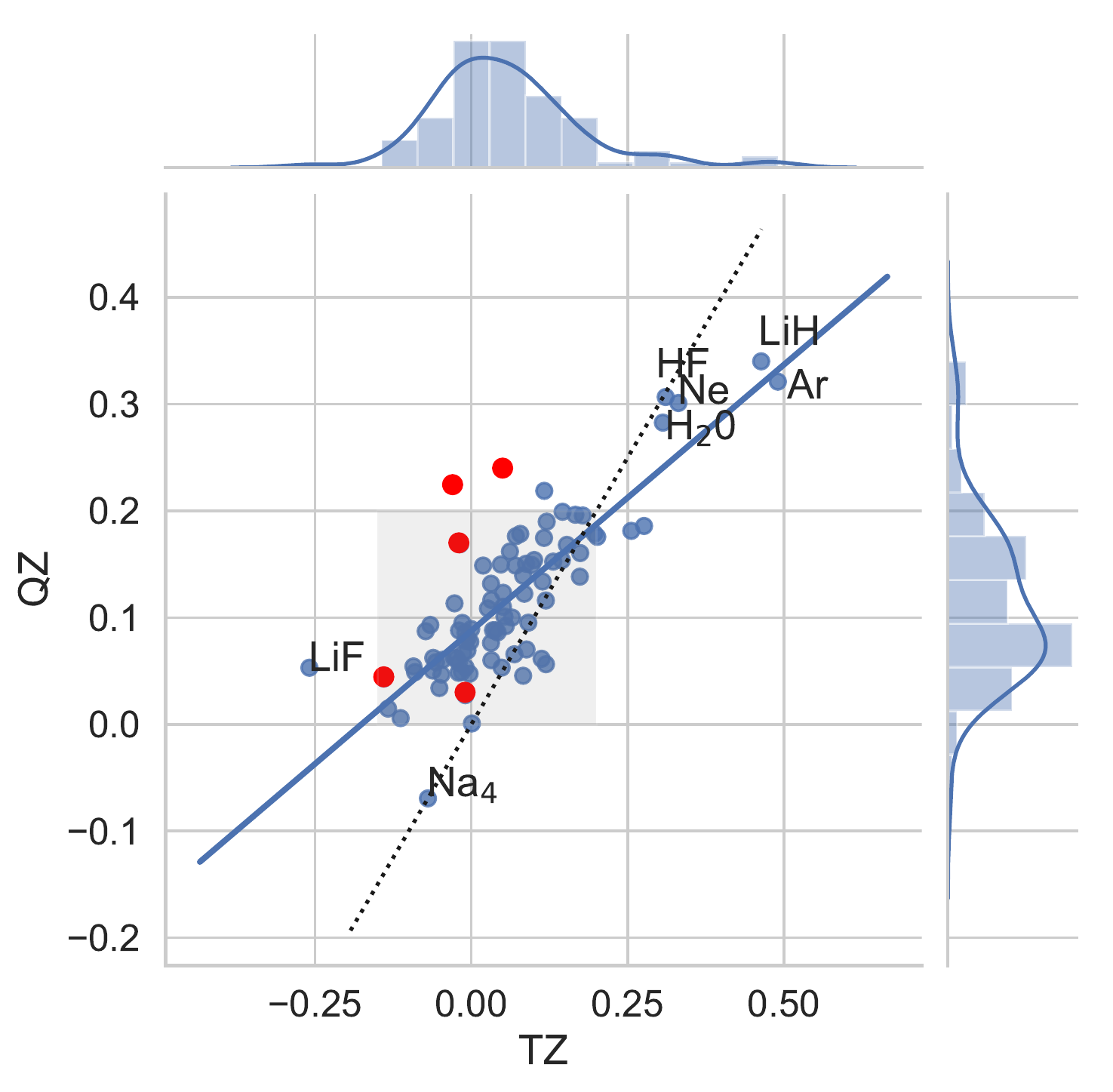}
    \caption{Distribution of deviations of the ADF IPs to the def2-GTO IPs for the GW100 database for TZ and QZ basis sets ($\Delta_{X,i}\;, X = \text{TZ,QZ}$). The univariate plots show the distribution of the $\Delta_{TZ,i}$ (upper histogram) and the $\Delta_{QZ,i}$ (left histogram). The bivariate plots shows the pairs $(\Delta_{TZ,i},\Delta_{QZ,i})$ and the blue line is a linear fit. The dotted line is defined by $\Delta_{T} = \Delta_{Q}$. Systems containing 5th row elements are highlighted in red. All values are in eV.}
    \label{fig::TMbse}
\end{figure}

Additional insight into the convergence properties of the STO type basis sets is provided when comparing them to GTO type ones of the same cardinality. Such a comparison is made in figure~\ref{fig::TMbse} for results on the TZ and and for the QZ level. The univariate plot on top of figure~\ref{fig::TMbse} shows the deviation of the aug-TZ3P IPs to the ones obtained with def2-TZVP and the univariate plot on the right side of  figure~\ref{fig::TMbse} shows the same for aug-QZ6P and def2-QZVP (here and in the remainder of this paper, we use the results calculated with the TURBOMOLE code whenever we refer to def2-GTO basis sets. We could have equally well used results obtained with other codes like FHI-AIMS or MOLGW). Again, Gaussian kernels are fitted to the univariate distributions. The bivariate plot shows the individual pairs $(\Delta_{T,i},\Delta_{Q,i})$, with $\Delta_{T,i}$ ($\Delta_{Q,i}$) being the differences between the IPs calculated with aug-TZ3P and def2-TZVP (aug-QZ6P and def2-QZVP) for the $i$th datapoint in the GW100 database. The dotted line is defined by $\Delta_{T,i} = \Delta_{Q,i}$. The systems for which the QP equations \eqref{quasi-particle-equations} can have multiple solutions (\ce{CI4}, \ce{KBr}, \ce{NaCl}, \ce{BN}, \ce{O3}, \ce{BeO}, \ce{MgO}, \ce{Cu2}, and \ce{CuCN}) are excluded from this comparison.\cite{Govoni2018} Also \ce{Ag2} is not shown since the deviations on the TZ and QZ level are exceptionally large.

We observe, that the maximum of the Gaussian kernel function is close to zero eV for the TZ and closer to 0.1 eV for the QZ basis sets. In other words, the aug-QZ6P IPs are consistently smaller than the def2-QZVP IPs (with a mean deviation (MD) of 120 eV), while the aug-TZ3P ones are with a MD of 60 meV only slightly smaller than the def2-TZVP IPs on average. The missing $l_{occ}+3$ function in the aug-QZ6P basis set might be a reason for this discrepancy. We will see in the next subsection that the CBS limit extrapolated IPs calculated with our basis sets are on average lower than the def2-GTO results. This section clearly shows that this discrepancy is mostly caused by the differences on the QZ level. 

The deviations of the STO-type to the respective def2- basis sets are strongly correlated. In cases in which the aug-TZ3P IPs are considerably smaller than their def2-TZVP counterparts, also the aug-QZ6P IPs will be much smaller than the def2-QZVP ones. Good examples are the five molecules represented by the points in the upper right corner of the bivariate plot. Since it is known that the GTO-type basis sets allow for a reliable CBS limit extrapolation, this fact is highly important since it guarantees that the same CBS limit extrapolation can be performed using our STO-type basis sets. 

We also shortly comment on the systems containing 5th-row elements, highlighted in red. With the exception of one of them (\ce{Rb2}), the agreement on the TZ is rather good, while they agreement on the QZ level is significantly worse. This is again due to the inconsistent polarization for the heavier elements mentioned above; approaching the CBS limit for heavier elements becomes difficult without Slater functions with angular momentum larger than $l=3$.

To summarize the key points of this section, our STO type basis sets seem to behave qualitatively similar to the GTO-type basis sets, although the improvement when going from TZ to QZ is smaller for the STO- than for the GTO-type basis sets. Together with the good correlation of deviations on the TZ and QZ level, this indicates that our basis sets allow indeed for a meaningful CBS limit extrapolation. However, the CBS limit extrapolated IPs from the STO-type basis sets will on average be lower than their counterparts calculated using GTOs.

\subsection{Comparison to Other Codes}

\subsubsection{KS Eigenvalues}
Before we dive into our comparison of the $GW$ QP energies, we shorty compare our KS eigenvalues for the systems in GW100 to the ones from other codes. Here and in the following, we do not include Phenol and Vynilbromide in the statistical analysis since different structures have been used in the past for both systems.\cite{Maggio2017a} Our results (see supporting information for the raw data for all 100 systems) only confirm what is already well known; For KS eigenvalues, the agreement between different codes is generally excellent. For example, the CBS limit extrapolated KS HOMO energies from the WEST code agree with the ones obtained from def-GTO calculations within 30 meV on average, with a maximum deviations of 176 meV.\cite{Govoni2018} These figures reduce to 24 meV and 92 meV when the plane-wave results are compared to def2-QZVP.\cite{Govoni2018} With only 19 meV, the agreement from VASP to def2-QZVP is even better.\cite{Maggio2017a} One should keep in mind that the extrapolation schemes for correlated-electron methods and localized basis functions are not necessarily useful to extrapolate KS eigenvalues as has already been pointed out in ref.~\citen{Maggio2017a} and in ref.~\citen{Govoni2018}. Such a comparison should rather be based on non-extrapolated results.\cite{Jensen2002,Kraus2020}

Our KS HOMO eigenvalues calculated on the aug-QZ6P level of theory show a MAD of 26 meV to the ones on the def2-QZVP level and of 22 meV to the CBS limit extrapolated values calculated with the WEST code. Our LUMO eigenvalues only differ to the ones from WEST by 35 meV on average. Major deviations of more than 150 meV are only found for Helium (340 meV), \ce{H2} (280 meV) and \ce{Ag2} (340 meV). With deviations of more than 420 meV to WEST and 490 meV to def-QZVP, the latter system is also the only outlier for IPs. However, when the ZORA is made, the deviations to WEST reduce to 15 meV for the HOMO and 17 meV for the LUMO, respectively. Also, the deviation to the def2-QZVP IP reduces to an acceptable value of 50 meV which is rather strange, given that the latter has been calculated without relativistic corrections.

\subsubsection{Ionization Potentials}

We now turn our attention to the IPs calculated within the GWA. In addition to Phenol and Vynilbromide, we again exclude the systems for which the QP equations \eqref{quasi-particle-equations} can have multiple solutions (\ce{CI4}, \ce{KBr}, \ce{NaCl}, \ce{BN}, \ce{O3}, \ce{BeO}, \ce{MgO}, \ce{Cu2}, and \ce{CuCN})\cite{Govoni2018} from the following statistical comparison, but also \ce{TiF4} and \ce{OCS} for which no IPs from the WEST code are available. Due to large discrepancies between WEST and VASP, we also exclude the systems containing Iodine, Gallium and Xenon. Finally, we also exclude all remaining systems containing 5th-row elements from our analysis since for these systems (especially the ones containing Iodine and \ce{Ag2}) the different treatment of relativistic effects have been shown to significantly affects QP energies.\cite{Maggio2017a} This leaves us with a set of 81 molecules whose IPs we include in the statistical analysis in this section. 

\renewcommand*{\arraystretch}{0.4}
\sisetup{
  round-mode          = places, 
  round-precision     = 2, 
}

\noindent\begin{longtable}[c]{ll
S[table-format=2.2]%
S[table-format=2.2]%
S[table-format=2.2]%
S[table-format=2.2]%
S[table-format=2.2]%
}
\caption{\label{tab::IPs}$G_0W_0$@PBE ionization potentials (IP) for the GW100 database (third column) Columns four to seven denote deviations of the ADF IPs to the ones from reference $X$, $\Delta_X = IP_{X} - IP_{ADF}$. All values are in eV.} \\
\toprule
& Name & {ADF} & {$\Delta_{def2-GTO}$} 
& {$\Delta_{VASP}$} & {$\Delta_{WEST}$} & {$\Delta_{nanoGW}$} \\
\midrule\endfirsthead\toprule 
& Name & {ADF} & {$\Delta_{def2-GTO}$} 
& {$\Delta_{VASP}$} & {$\Delta_{WEST}$} & {$\Delta_{nanoGW}$} \\
\midrule\endhead\bottomrule\midrule%
\multicolumn{7}{r}{{Continued on next page}} \\ \bottomrule
\endfoot\bottomrule\endlastfoot
   1 &               Helium &  23.31 &   0.18 &   0.07 &   0.11 &  -0.11 \\
   2 &                 Neon &  20.06 &   0.27 &   0.11 &   0.27 &   0.39 \\
   3 &                Argon &  15.26 &   0.02 &   0.06 &   0.11 &   0.19 \\
   4 &              Krypton &  13.71 &   0.18 &   0.22 &   0.05 &  -0.06 \\
   5 &                Xenon &  11.88 &   0.45 &   0.26 &   1.24 &   0.23 \\
   6 &             Hydrogen &  15.88 &  -0.03 &  -0.03 &  -0.04 &  -0.13 \\
   7 &        Lithium dimer &   5.08 &  -0.02 &   0.01 &  -0.04 &  -0.04 \\
   8 &         Sodium dimer &   4.89 &   0.02 &   0.04 &   0.09 &   0.05 \\
   9 &      Sodium tetramer &   4.25 &  -0.04 &  -0.08 &  -0.01 &  -0.03 \\
  10 &       Sodium hexamer &   4.30 &   0.04 &   0.04 &   0.07 &   0.05 \\
  11 &      Potassium dimer &   3.99 &   0.09 &   0.13 &   0.15 &   0.07 \\
  12 &       Rubidium dimer &   3.78 &   0.10 &   0.24 &   0.23 &   0.08 \\
  13 &             Nitrogen &  14.79 &   0.26 &   0.14 &   0.15 &   0.04 \\
  14 &     Phosphorus dimer &  10.26 &   0.12 &   0.09 &   0.17 &   0.06 \\
  15 &        Arsenic dimer &   9.58 &   0.08 &   0.01 &  -0.03 &  -0.09 \\
  16 &             Fluorine &  14.99 &   0.12 &  -0.06 &   0.01 &   0.03 \\
  17 &             Chlorine &  11.18 &   0.13 &   0.14 &   0.23 &   0.13 \\
  18 &              Bromine &  10.46 &   0.10 &   0.11 &  -0.02 &  -0.17 \\
  19 &               Iodine &   9.04 &   0.54 &   0.48 &   1.37 &   0.23 \\
  20 &              Methane &  13.89 &   0.11 &   0.13 &   0.10 &   0.26 \\
  21 &               Ethane &  12.35 &   0.11 &   0.15 &   0.09 &  -0.19 \\
  22 &              Propane &  11.85 &   0.04 &   0.05 &  -0.01 &  -0.32 \\
  23 &               Butane &  11.52 &   0.07 &   0.09 &  -0.11 &  -0.27 \\
  24 &             Ethylene &  10.28 &   0.12 &   0.14 &   0.11 &  -0.01 \\
  25 &            Acetylene &  11.14 &  -0.05 &  -0.07 &  -0.05 &  -0.21 \\
  26 &          Tetracarbon &  10.74 &   0.17 &   0.15 &   0.16 &   0.06 \\
  27 &         Cyclopropane &  10.63 &   0.02 &   0.09 &   0.04 &  -0.15 \\
  28 &              Benzene &   9.07 &   0.03 &   0.04 &   0.01 &  -0.11 \\
  29 &    Cyclooctatetraene &   8.16 &   0.02 &   0.03 &   0.00 &  -0.14 \\
  30 &      Cyclopentadiene &   8.43 &   0.02 &   0.04 &   0.01 &  -0.13 \\
  31 &       Vinyl fluoride &  10.24 &   0.08 &   0.04 &   0.05 &  -0.06 \\
  32 &       Vinyl chloride &   9.82 &   0.08 &   0.10 &   0.12 &  -0.01 \\
  33 &        Vinyl bromide &   9.03 &   0.11 &   0.72 &   0.61 &   0.49 \\
  34 &         Vinyl iodide &   8.93 &   0.26 &   0.34 &   0.88 &   0.09 \\
  35 &   Tetrafluoromethane &  15.43 &   0.17 &  -0.02 &   0.08 &  -0.04 \\
  36 &   Tetrachloromethane &  11.24 &  -0.03 &  -0.04 &   0.05 &  -0.00 \\
  37 &    Tetrabromomethane &  10.05 &   0.17 &   0.20 &   0.06 &  -0.01 \\
  38 &     Tetraiodomethane &   8.74 &   0.31 &   0.37 &        &   0.07 \\
  39 &               Silane &  12.37 &   0.02 &   0.03 &   0.05 &  -0.12 \\
  40 &              Germane &  12.09 &   0.03 &   0.04 &   0.23 &  -0.18 \\
  41 &             Disilane &  10.46 &  -0.05 &  -0.02 &   0.06 &  -0.14 \\
  42 &          Pentasilane &   9.07 &  -0.02 &   0.06 &   0.12 &  -0.08 \\
  43 &      Lithium hydride &   6.52 &   0.07 &  -0.06 &   0.11 &   0.15 \\
  44 &    Potassium hydride &   4.88 &   0.09 &   0.09 &   0.09 &  -0.20 \\
  45 &               Borane &  12.93 &   0.03 &   0.02 &   0.02 &  -0.15 \\
  46 &          Diborane(6) &  11.92 &   0.01 &   0.02 &  -0.00 &  -0.25 \\
  47 &               Amonia &  10.27 &   0.12 &   0.05 &  -0.09 &  -0.20 \\
  48 &       Hydrazoic acid &  10.44 &   0.11 &   0.06 &   0.04 &  -0.26 \\
  49 &            Phosphine &  10.27 &   0.08 &   0.08 &   0.16 &   0.04 \\
  50 &               Arsine &  10.31 &  -0.11 &  -0.05 &   0.02 &  -0.11 \\
  51 &     Hydrogen sulfide &  10.12 &   0.01 &  -0.01 &   0.11 &   0.07 \\
  52 &    Hydrogen fluoride &  15.14 &   0.23 &   0.23 &   0.09 &   0.12 \\
  53 &    Hydrogen chloride &  12.36 &   0.00 &   0.09 &   0.12 &   0.11 \\
  54 &     Lithium fluoride &  10.03 &   0.24 &   0.04 &   0.08 &   0.03 \\
  55 &   Magnesium fluoride &  12.44 &   0.07 &  -0.03 &   0.02 &  -0.14 \\
  56 &Titanium tetrafluoride&  13.93 &   0.15 &   0.08 &        &   0.00 \\
  57 &    Aluminum fluoride &  14.33 &   0.15 &   0.00 &   0.07 &  -0.13 \\
  58 &   Boron monofluoride &  10.59 &   0.14 &  -0.13 &  -0.03 &  -0.17 \\
  59 & Sulfur tetrafluoride &  12.34 &   0.05 &  -0.14 &  -0.02 &  -0.18 \\
  60 &    Potassium bromide &   7.85 &  -0.24 &  -0.05 &        &  -0.74 \\
  61 & Gallium monochloride &   9.81 &  -0.07 &   0.08 &   0.36 &  -0.06 \\
  62 &      Sodium chloride &   8.32 &   0.03 &   0.15 &        &  -0.14 \\
  63 &   Magnesium chloride &  11.07 &   0.12 &   0.12 &   0.18 &   0.04 \\
  64 &      Aluminum iodide &   9.38 &   0.20 &   0.20 &   0.93 &  -0.07 \\
  65 &        Boron nitride &  10.94 &   0.15 &        &        &   0.25 \\
  66 &     Hydrogen cyanide &  13.10 &   0.22 &   0.19 &   0.12 &   0.12 \\
  67 &Phosphorus mononitrid &  11.10 &   0.20 &   0.14 &   0.16 &   0.11 \\
  68 &            Hydrazine &   9.40 &  -0.03 &  -0.07 &  -0.13 &  -0.30 \\
  69 &         Formaldehyde &  10.46 &   0.00 &  -0.04 &  -0.05 &  -0.25 \\
  70 &             Methanol &  10.54 &   0.13 &   0.07 &   0.07 &  -0.23 \\
  71 &              Ethanol &  10.17 &   0.10 &   0.04 &   0.04 &  -0.20 \\
  72 &         Acetaldehyde &   9.61 &   0.05 &   0.02 &   0.00 &  -0.22 \\
  73 &        Ethoxy ethane &   9.41 &   0.02 &   0.02 &  -0.02 &  -0.30 \\
  74 &          Formic acid &  10.84 &   0.03 &  -0.03 &  -0.03 &  -0.13 \\
  75 &    Hydrogen peroxide &  10.96 &   0.14 &  -0.00 &   0.04 &   0.02 \\
  76 &                Water &  11.94 &   0.11 &  -0.10 &  -0.07 &  -0.12 \\
  77 &       Carbon dioxide &  13.37 &   0.09 &  -0.01 &   0.00 &  -0.25 \\
  78 &     Carbon disulfide &   9.80 &   0.15 &   0.16 &   0.25 &   0.09 \\
  79 & Carbon oxide sulfide &  10.90 &   0.21 &   0.16 &   0.26 &   0.08 \\
  80 &Carbon oxide selenide &  10.40 &   0.03 &   0.02 &  -0.03 &  -0.22 \\
  81 &      Carbon monoxide &  13.66 &   0.05 &  -0.04 &  -0.00 &  -0.18 \\
  82 &                Ozone &  11.74 &  -0.23 &        &        &   0.33 \\
  83 &       Sulfur dioxide &  11.86 &   0.20 &   0.05 &   0.10 &   0.05 \\
  84 &   Beryllium monoxide &   8.99 &  -0.48 &        &        &   0.76 \\
  85 &   Magnesium monoxide &   6.82 &  -0.07 &        &        &   0.24 \\
  86 &              Toluene &   8.72 &   0.00 &   0.03 &  -0.01 &  -0.14 \\
  87 &         Ethylbenzene &   8.65 &   0.01 &   0.04 &   0.01 &  -0.16 \\
  88 &    Hexafluorobenzene &   9.70 &   0.04 &  -0.07 &  -0.05 &  -0.13 \\
  89 &               Phenol &   8.47 &   0.04 &  -0.09 &        &  -0.24 \\
  90 &              Aniline &   7.72 &   0.07 &   0.06 &   0.01 &  -0.14 \\
  91 &             Pyridine &   9.18 &   0.00 &  -0.02 &  -0.05 &  -0.19 \\
  92 &              Guanine &   7.77 &   0.10 &   0.08 &   0.05 &  -0.12 \\
  93 &              Adenine &   8.14 &   0.01 &  -0.02 &  -0.05 &  -0.20 \\
  94 &             Cytosine &   8.44 &   0.01 &  -0.04 &  -0.04 &  -0.19 \\
  95 &              Thymine &   8.86 &   0.01 &  -0.03 &  -0.04 &  -0.22 \\
  96 &               Uracil &   9.26 &   0.12 &   0.10 &  -0.07 &  -0.02 \\
  97 &                 Urea &   9.25 &   0.21 &   0.10 &   0.14 &  -0.03 \\
  98 &         Silver dimer &   7.06 &   0.91 &   0.77 &   0.98 &   0.81 \\
  99 &         Copper dimer &   8.20 &  -1.57 &  -1.01 &        &  -0.33 \\
 100 &       Copper cyanide &  10.25 &  -0.73 &        &        &   0.40 \\
\bottomrule 
\end{longtable} 

\begin{figure}[ht]
    \centering
    \includegraphics[width=0.8\textwidth]{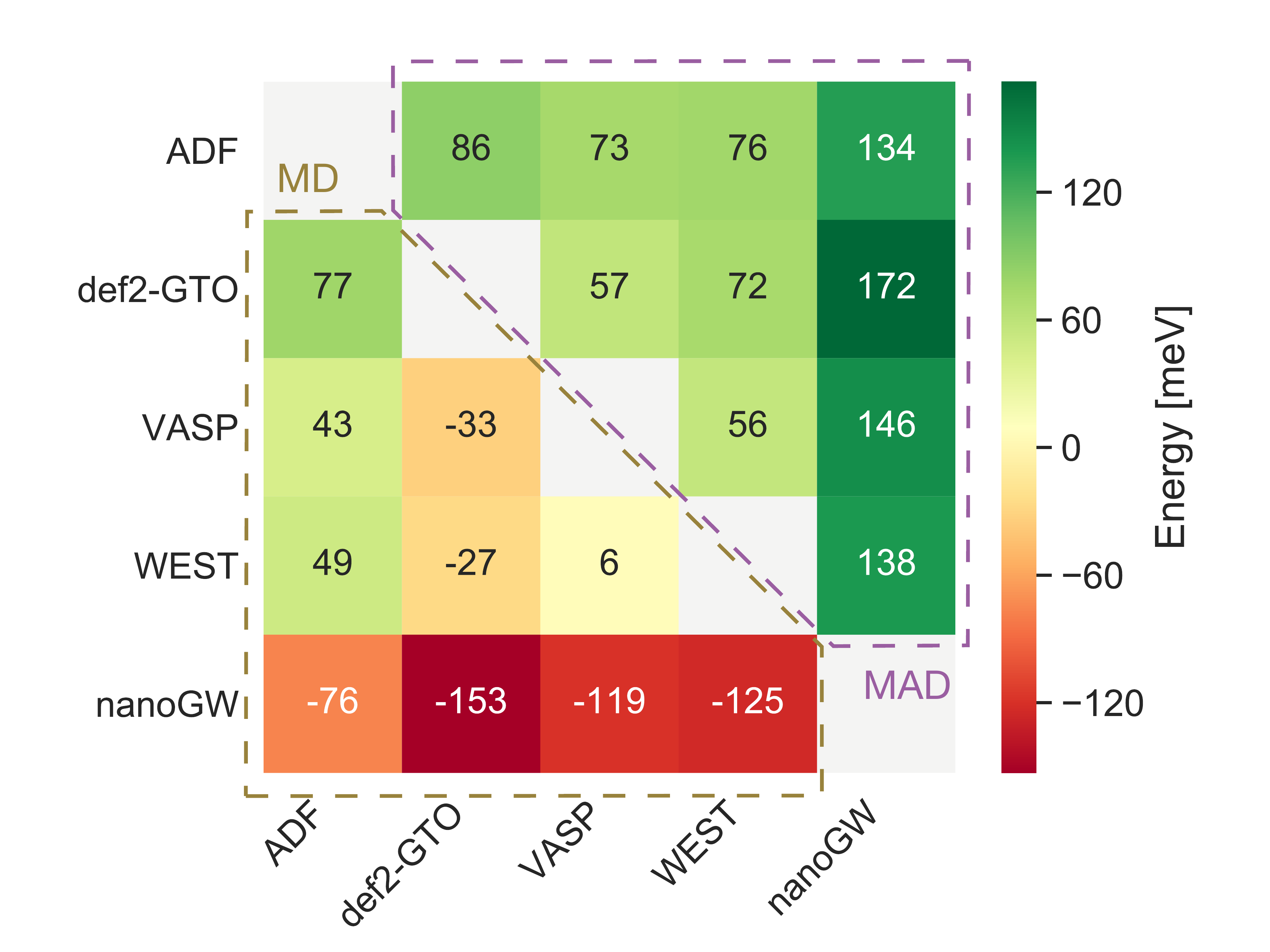}
    \caption{MADs, (upper triangle) and mean deviations (MD) (lower triangle) of the CBS limit extrapolated IPs in the GW100 database (18 molecules have been excluded from the comparison, see explanations above) computed with different codes. All values are in meV.}
    \label{fig::MADs_HOMO}
\end{figure}

\begin{figure}[ht]
    \centering
    \includegraphics[width=0.7\textwidth]{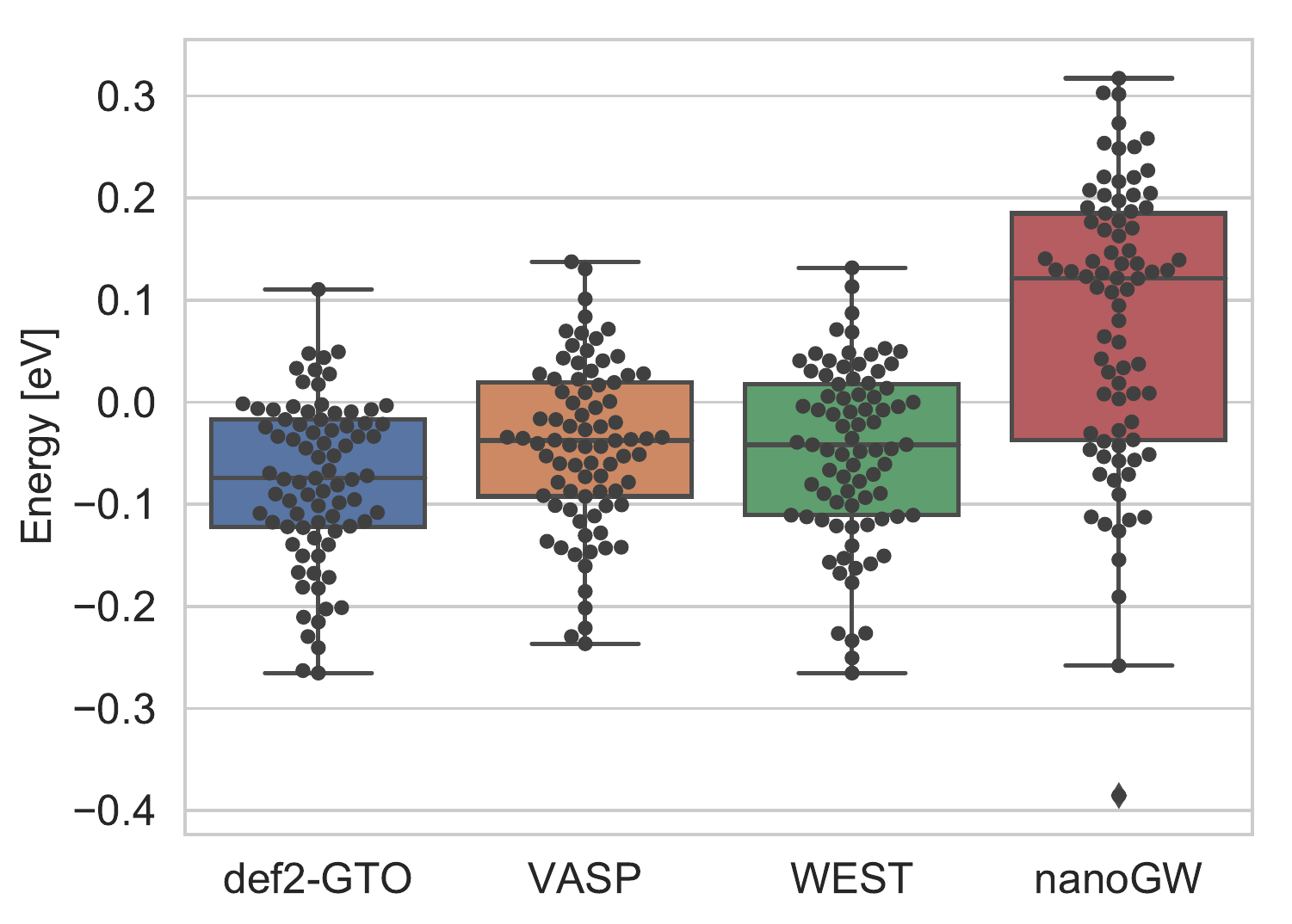}
    \caption{Deviation of IPs calculated with ADF to different codes. Black dots denote the individual data points. The horizonal line in each box denotes the median deviation, the box contains all data points between the first quartile (Q1) and third quartile (Q2) and the whiskers are at $Q1 \pm \frac{3}{2}|Q1-Q3|$ (in case of a normal distribution, the whiskers include 99.3 \% of all data points). All values are in eV.}
    \label{fig::boxplot_HOMO}
\end{figure}

Table~\ref{tab::IPs} shows the results for all 100 IPs in the GW100 database obtained with our code next to the deviations to def2-GTO, VASP, WEST and nanoGW, if available. To facilitate a discussion of the data, figures~\ref{fig::MADs_HOMO} shows MADs and maximum absolute errors (MAE) between all codes, while figure~\ref{fig::boxplot_HOMO} visualizes the distribution of the deviations of the ADF IPs to the ones from other codes. The IPs from ADF, def2-GTO, VASP and WEST are all in good agreement with each other, with MADs between 56 and 86 meV, while the deviations to nanoGW are about twice as large. Figure~\ref{fig::boxplot_HOMO} also shows, that the deviations of the ADF IPs to the ones from other codes (again, with the exception of nanoGW) show a small spread and no outliers can be found. We note again, that we assume the CBS limit extrapolation error to be of the order of at least 60 meV on average and all the values reported and compared here should only be interpreted with these error bars. For the plane-wave codes, the CBS limit extrapolation error is likely smaller than for the localized basis sets, but there are additional sources of error, most notably pseudo-potentials and box-size effects. In light of these uncertainties, the agreement between all four codes can be considered as excellent.

Looking at the mean deviations to the other codes in the lower triangle of figure~\ref{fig::MADs_HOMO} as well as at the boxplots in figure~\ref{fig::boxplot_HOMO}, we see that ADF IPs are generally smaller than the ones from def2-GTO, VASP and WEST. The reasons for the discrepancy between ADF and def2-GTO have already been discussed above. We also see that the nanoGW IPs are on average much smaller than the ones from all others codes. This is in line with the fact that the nanoGW results were apparently obtained without basis set extrapolation,\cite{Gao2019} although the numerical parameters determining the convergence to the CBS limit (grid spacing, chosen cut-off for virtual states and radius of the sphere around a given finite systems) where tested for convergence separately. Still, our data analysis suggests that the nanoGW IPs are not as well converged as the ones from the other codes.

Systems containing Fluorine and Nitrogen generally show rather pronounced disagreements between different codes (For example, consider the following deviations from ADF to def2-GTO: \ce{N2}: 260 meV, \ce{HCN}: 220 meV, \ce{TiF4}: 150 meV, \ce{AlF3}: 150 meV, \ce{LiF}: 240 meV, \ce{HF}: 230 meV). For Fluorine, this has already been observed by Maggio and Kresse in ref.~\citen{Maggio2017a}, who pointed at the default pseudo-potentials in VASP as a potential source of these discrepancies. Another possible explanation might be found in a a recent study by Bruneval et al.\cite{Bruneval2020} which suggests, that molecules predominantly composed of Carbon and Hydrogen converge to the CBS limit rather quickly, whereas the convergence is considerably slower for systems to a large part composed of Fluorine and Nitrogen. For this reason, we also used stronger polarized basis sets for Fluorine and Chlorine (For the latter, the basis set convergence is also rather slow, but less pronounced than for Fluorine).

Finally, we shortly comment on some systems which we have excluded from the statistical comparisons in figures~\ref{fig::MADs_HOMO} and figures~\ref{fig::boxplot_HOMO}. We find large differences for system containing 5th row elements, e.g. \ce{Xe}, ce{Ag2}, \ce{I2}, or \ce{CI4}. Here, the ADF IPs are considerably lower than the ones from other codes, indicating that the ADF results are not properly converged to the CBS limit which is due to the missing basis functions with angular momentum higher than $l=3$. 

With 1.57 eV, the by far largest deviation reported in table~\ref{tab::IPs} can be found between ADF (8.20 eV) and TURBOMOLE (6.63 eV) for \ce{Cu2}. FHI-AIMS gives an IP of 7.78 eV for this system,\cite{VanSetten2015} which is in considerably better agreement with ADF. FHI-AIMS relies on an analytical continuation from the imaginary to the real frequency axis with 16 sampling points (AIMS-P16), not much different from the procedure in ADF. We can conclude, that for this particular system, the large deviation of ADF and FHI-AIMS to TURBOMOLE are caused by inaccuracies in the frequency grids, which are not converged. The ADF IP reported in table~\ref{tab::IPs} has been obtained from aug-TZ3P and aug-QZ6P calculation with 24 and 27 imaginary frequency points, respectively. When 24 imaginary frequency points are used for the aug-QZ6P calculation as well, the IP of \ce{Cu2} reduces to 8.05 eV, which is already in reasonable agreement with FHI-AIMS. Similar conclusions can be drawn for \ce{CuCN}. Another interesting case is \ce{BeO}, with deviation of 0.48 eV to def-GTO. Again, the large deviation is due to non-converged frequency grids. the ADF IP is with 8.99 eV very close to the 9.07 eV obtained by AIMS-P16. With 128 imaginary frequency points (AIMS-P128), FHI-AIMS gives an IP of 9.63 eV which is then in perfect agreement with TURBOMOLE. Furthermore, in ref.~\citen{VanSetten2015}, three solutions are reported for \ce{BeO}, while ADF only recovers one of them. These three examples show, that the current frequency treatment in ADF can not properly describe the IPs of systems for which the single QP picture breaks down in the valence region.

\subsubsection{Electron Affinities}

We now turn our attention to the EAs. As for the IPs, table~\ref{tab::EAs} shows the EAs calculated with ADF and the differences to the other four codes excluding all noble gases and \ce{H2}. However, it is known, that the def2-GTO basis sets sometimes severely overestimate positive LUMO energies which then deviate from results from plane-wave codes by more than 1 eV. Furthermore, since EAs converge slower to the CBS limit than IPs when non-augmented basis sets are used,\cite{Bruneval2020, Forster2020b} also basis set extrapolation errors are larger for the remaining systems. On the other hand, plane-wave calculations require very large box sizes for these systems which makes it harder to converge the EAs with respect to this parameter and for this reason results from VASP are often not available.\cite{Maggio2017a} 

\renewcommand*{\arraystretch}{0.4}
\sisetup{
  round-mode          = places, 
  round-precision     = 2, 
}
\noindent\begin{longtable}[c]{ll
S[table-format=2.2]%
S[table-format=2.2]%
S[table-format=2.2]%
S[table-format=2.2]%
S[table-format=2.2]%
}
\caption{\label{tab::EAs}$G_0W_0$@PBE electron affinities (EA) for the GW100 database (third column) Columns four to seven denote deviations of the ADF EAs to the ones from reference $X$, $\Delta_X = EA_{X} - EA_{ADF}$. All values are in eV.}\\
\toprule
& Name & {ADF} & {$\Delta_{def2-GTO}$} 
& {$\Delta_{VASP}$} & {$\Delta_{WEST}$} & {$\Delta_{nanoGW}$} \\
\midrule\endfirsthead\toprule 
& Name & {ADF} & {$\Delta_{def2-GTO}$} 
& {$\Delta_{VASP}$} & {$\Delta_{WEST}$} & {$\Delta_{nanoGW}$} \\
\midrule\endhead\bottomrule\midrule%
\multicolumn{7}{r}{{Continued on next page}} \\ \bottomrule
\endfoot\bottomrule\endlastfoot\\
   7 &        Lithium dimer &   0.52 &   0.23 &   0.09 &   0.12 &   0.14 \\
   8 &         Sodium dimer &   0.64 &   0.02 &  -0.04 &  -0.03 &  -0.00 \\
   9 &      Sodium tetramer &   0.92 &   0.23 &   0.15 &   0.16 &   0.20 \\
  10 &       Sodium hexamer &   0.95 &   0.18 &   0.12 &   0.09 &   0.11 \\
  11 &      Potassium dimer &   0.59 &   0.16 &   0.15 &   0.16 &   0.16 \\
  12 &       Rubidium dimer &   0.67 &        &   0.07 &   0.06 &   0.07 \\
  13 &             Nitrogen &  -2.40 &   0.28 &        &   0.25 &   0.17 \\
  14 &     Phosphorus dimer &   0.64 &   0.44 &   0.35 &   0.45 &   0.38 \\
  15 &        Arsenic dimer &   1.08 &   0.44 &  -0.01 &   0.01 &  -0.04 \\
  16 &             Fluorine &   0.54 &   0.69 &        &   0.52 &   0.58 \\
  17 &             Chlorine &   0.83 &   0.57 &   0.42 &   0.55 &   0.52 \\
  18 &              Bromine &   1.40 &   0.56 &   0.59 &   0.48 &   0.29 \\
  19 &               Iodine &   1.56 &        &   0.65 &   1.65 &        \\
  20 &              Methane &  -0.78 &  -1.25 &   0.15 &   0.02 &  -0.10 \\
  21 &               Ethane &  -0.77 &  -1.16 &        &  -0.01 &  -0.11 \\
  22 &              Propane &  -0.72 &  -1.15 &        &  -0.03 &  -0.11 \\
  23 &               Butane &  -0.70 &  -1.13 &        &  -0.04 &  -0.13 \\
  24 &             Ethylene &  -1.91 &   0.09 &        &   0.11 &  -0.04 \\
  25 &            Acetylene &  -2.48 &  -0.08 &        &  -0.02 &  -0.20 \\
  26 &          Tetracarbon &   2.62 &   0.53 &   0.47 &   0.48 &   0.52 \\
  27 &         Cyclopropane &  -0.73 &  -1.23 &        &  -0.02 &  -0.14 \\
  28 &              Benzene &  -0.96 &   0.07 &        &   0.03 &  -0.10 \\
  29 &    Cyclooctatetraene &   0.03 &   0.09 &   0.02 &   0.04 &  -0.10 \\
  30 &      Cyclopentadiene &  -0.91 &   0.06 &        &   0.01 &  -0.15 \\
  31 &       Vinyl fluoride &  -1.92 &   0.04 &        &   0.03 &  -0.09 \\
  32 &       Vinyl chloride &  -1.31 &   0.14 &   0.12 &   0.10 &   0.00 \\
  33 &        Vinyl bromide &  -1.23 &   0.12 &        &   0.17 &   0.09 \\
  34 &         Vinyl iodide &  -0.77 &        &   0.40 &   0.55 &   0.44 \\
  35 &   Tetrafluoromethane &  -0.88 &  -3.00 &        &   0.05 &  -0.10 \\
  36 &   Tetrachloromethane &   0.04 &   0.50 &   0.28 &   0.37 &   0.41 \\
  37 &    Tetrabromomethane &   0.99 &   0.57 &   0.48 &   0.46 &   0.41 \\
  38 &     Tetraiodomethane &   2.16 &        &   0.28 &   0.88 &   0.32 \\
  39 &               Silane &  -0.72 &  -1.54 &        &  -0.04 &  -0.11 \\
  40 &              Germane &  -0.47 &  -1.38 &        &  -0.14 &  -0.36 \\
  41 &             Disilane &  -0.75 &  -0.76 &        &  -0.02 &  -0.83 \\
  42 &          Pentasilane &  -0.08 &   0.08 &   0.05 &   0.15 &  -0.02 \\
  43 &      Lithium hydride &   0.05 &   0.11 &   0.02 &   0.02 &   0.13 \\
  44 &    Potassium hydride &   0.17 &   0.15 &   0.08 &   0.08 &   0.35 \\
  45 &               Borane &  -0.26 &   0.23 &   0.23 &   0.25 &   0.21 \\
  46 &          Diborane(6) &  -0.87 &   0.13 &        &   0.15 &  -0.00 \\
  47 &               Amonia &  -0.76 &  -1.24 &        &  -0.05 &  -0.09 \\
  48 &       Hydrazoic acid &  -1.40 &   0.30 &        &   0.25 &   0.23 \\
  49 &            Phosphine &  -0.67 &  -1.59 &        &  -0.03 &  -0.06 \\
  50 &               Arsine &  -0.58 &  -1.36 &        &  -0.08 &  -0.16 \\
  51 &     Hydrogen sulfide &  -0.73 &  -1.52 &        &  -0.05 &  -0.11 \\
  52 &    Hydrogen fluoride &  -1.06 &  -0.98 &        &  -0.05 &  -0.12 \\
  53 &    Hydrogen chloride &  -1.19 &  -0.34 &        &   0.10 &  -0.03 \\
  54 &     Lithium fluoride &  -0.04 &   0.05 &  -0.13 &  -0.03 &   0.11 \\
  55 &   Magnesium fluoride &   0.26 &   0.05 &   0.03 &   0.07 &   0.18 \\
  56 & itanium tetrafluorid &   0.09 &   0.97 &        &   0.83 &   1.14 \\
  57 &    Aluminum fluoride &   0.06 &   0.17 &  -0.14 &   0.10 &   0.01 \\
  58 &   Boron monofluoride &  -1.21 &   0.16 &        &   0.28 &   0.24 \\
  59 & Sulfur tetrafluoride &  -0.29 &   0.39 &   0.22 &   0.37 &   0.35 \\
  60 &    Potassium bromide &   0.34 &   0.08 &  -0.02 &   0.06 &   0.59 \\
  61 & Gallium monochloride &   0.02 &   0.37 &   0.17 &   0.42 &   0.20 \\
  62 &      Sodium chloride &   0.42 &  -0.00 &   0.04 &   0.05 &   0.61 \\
  63 &   Magnesium chloride &   0.68 &  -0.00 &  -0.07 &   0.02 &   0.05 \\
  64 &      Aluminum iodide &   1.18 &        &  -0.16 &   0.48 &   0.02 \\
  65 &        Boron nitride &   4.05 &  -0.10 &        &   0.03 &   0.05 \\
  66 &     Hydrogen cyanide &  -2.31 &   0.09 &        &   0.06 &  -0.06 \\
  67 & hosphorus mononitrid &   0.12 &   0.47 &        &   0.40 &   0.35 \\
  68 &            Hydrazine &  -0.70 &  -0.98 &        &  -0.02 &  -0.08 \\
  69 &         Formaldehyde &  -1.06 &   0.35 &        &   0.30 &   0.15 \\
  70 &             Methanol &  -0.81 &  -1.00 &        &  -0.10 &  -0.19 \\
  71 &              Ethanol &  -0.73 &  -0.94 &        &  -0.11 &  -0.19 \\
  72 &         Acetaldehyde &  -1.16 &   0.33 &   0.29 &   0.29 &   0.25 \\
  73 &        Ethoxy ethane &  -0.62 &  -1.08 &        &  -0.09 &  -0.18 \\
  74 &          Formic acid &  -1.82 &   0.23 &   0.18 &   0.18 &   0.04 \\
  75 &    Hydrogen peroxide &  -2.06 &   0.11 &        &   0.26 &   0.28 \\
  76 &                Water &  -0.88 &  -1.13 &  -0.16 &  -0.03 &  -0.08 \\
  77 &       Carbon dioxide &  -1.03 &   0.10 &        &   0.06 &  -0.03 \\
  78 &     Carbon disulfide &   0.10 &   0.45 &   0.32 &   0.40 &   0.32 \\
  79 & Carbon oxide sulfide &  -1.22 &   0.39 &        &   0.28 &   0.20 \\
  80 & arbon oxide selenide &  -0.93 &   0.41 &        &   0.29 &   0.22 \\
  81 &      Carbon monoxide &  -0.84 &   0.47 &        &   0.40 &   0.38 \\
  82 &                Ozone &   2.03 &   0.66 &   0.47 &   0.53 &   0.52 \\
  83 &       Sulfur dioxide &   0.86 &   0.63 &   0.39 &   0.51 &   0.36 \\
  84 &   Beryllium monoxide &   1.99 &   0.73 &   0.74 &   0.52 &   0.54 \\
  85 &   Magnesium monoxide &   1.74 &   0.39 &   0.31 &   0.21 &   0.52 \\
  86 &              Toluene &  -0.91 &   0.08 &        &   0.04 &  -0.09 \\
  87 &         Ethylbenzene &  -0.87 &   0.00 &        &  -0.03 &  -0.18 \\
  88 &    Hexafluorobenzene &  -0.03 &  -0.33 &        &   0.00 &  -0.08 \\
  89 &               Phenol &  -0.78 &   0.04 &        &  -0.07 &  -0.18 \\
  90 &              Aniline &  -0.91 &  -0.03 &        &  -0.07 &  -0.19 \\
  91 &             Pyridine &  -0.44 &   0.14 &        &   0.08 &  -0.05 \\
  92 &              Guanine &  -0.48 &   0.02 &        &  -0.03 &  -0.15 \\
  93 &              Adenine &  -0.28 &   0.07 &        &  -0.01 &  -0.13 \\
  94 &             Cytosine &  -0.18 &   0.17 &   0.06 &   0.09 &  -0.03 \\
  95 &              Thymine &   0.02 &   0.16 &   0.04 &   0.08 &  -0.06 \\
  96 &               Uracil &   0.05 &   0.20 &   0.06 &   0.10 &  -0.02 \\
  97 &                 Urea &  -0.49 &  -0.68 &        &  -0.04 &  -0.12 \\
  98 &         Silver dimer &   0.91 &        &   0.44 &   0.58 &   0.41 \\
  99 &         Copper dimer &   1.00 &   0.23 &   0.24 &   0.41 &   0.25 \\
 100 &       Copper cyanide &   1.47 &   0.38 &   0.44 &   0.51 &   0.34 \\
 MAD &                      &        &   0.48 &   0.21 &  0.16  &   0.21 \\ 
\bottomrule 
\end{longtable} 

\begin{figure}[ht]
    \centering
    \includegraphics[width=0.7\textwidth]{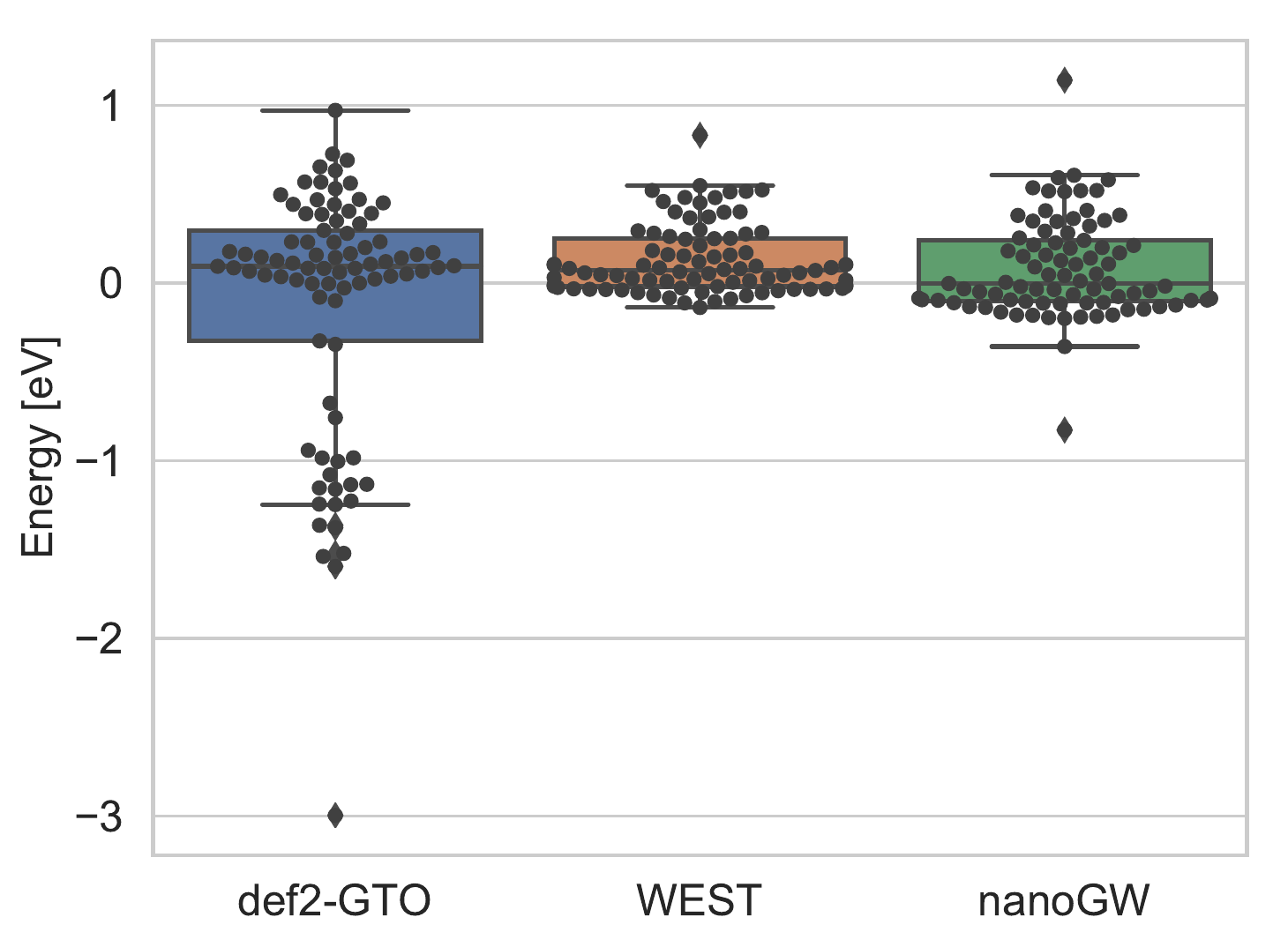}
    \caption{Deviation of EAs calculated with ADF to different codes. Black dots denote the individual data points. The horizonal line in each box denotes the median deviation, the box contains all data points between the first quartile (Q1) and third quartile (Q2) and the whiskers are at $Q1 \pm \frac{3}{2}|Q1-Q3|$ (in case of a normal distribution, the whiskers include 99.3 \% of all data points). All values are in eV. The deviations to VASP have been excluded due to the lack of reference values for too many systems in GW100.}
    \label{fig::boxplot_LUMO}
\end{figure}

Thus, for the full database, only comparison to WEST and nanoGW is possible. Excluding again all compounds containing Iodine, Copper, Gallium and Xenon as well as remaining systems containing 5th row elements, we find a MAD of 160 meV to the former and of 210 meV to the latter code. These MADs are about twice as large as for IPs but in light of the difficulties mentioned above certainly not surprising and in line with the previous benchmark studies on GW100.\cite{VanSetten2015, Maggio2017a, Govoni2018, Gao2019} Figure~\ref{fig::boxplot_LUMO} shows that the ADF EAs, as for the IPs, are on average smaller than the ones from WEST while no trend in that direction can be observed when comparing to nanoGW.

\begin{table}
    \centering
    \begin{tabular}{llccccccccccc}
    \toprule 
                 & & \multicolumn{4}{c}{aug.} 
                 &  \multicolumn{4}{c}{non.aug.} & \\
                 \cline{3-6}
                 \cline{7-10}
                 Index &  & T & Q & Ex. & $\Delta_{TQ}$ & T & Q & Ex. & $\Delta_{TQ}$ 
                 & $\Delta_{TT}$ & $\Delta_{QQ}$& $\Delta_{EE}$ \\
    \midrule
    13 & \ce{N2}    & -2.65 & -2.54 & -2.40 & -0.11 & -3.00 & -2.58 & -2.15 & -0.42 & 0.36 & 0.04 & -0.25 \\
    20 & \ce{CH4}   & -0.97 & -0.89 & -0.78 & -0.08 & -2.20 & -1.59 & -0.95 & -0.61 & 1.23 & 0.70 & 0.17  \\
    21 & \ce{C2H6}  & -0.96 & -0.88 & -0.77 & -0.08 & -2.13 & -1.52 & -0.87 & -0.62 & 1.17 & 0.64 & 0.10  \\
    22 & \ce{C3H8}  & -0.92 & -0.83 & -0.72 & -0.08 & -2.06 & -1.44 & -0.78 & -0.62 & 1.14 & 0.61 & 0.06  \\
    23 & \ce{C4H10} & -0.89 & -0.81 & -0.70 & -0.08 & -2.04 & -1.41 & -0.74 & -0.63 & 1.15 & 0.60 & 0.04  \\
    24 & \ce{C2H4}  & -2.12 & -2.03 & -1.91 & -0.09 & -2.40 & -2.12 & -1.81 & -0.28 & 0.28 & 0.09 & -0.10 \\
    25 & \ce{C2H2}  & -2.76 & -2.65 & -2.48 & -0.11 & -3.24 & -2.87 & -2.44 & -0.37 & 0.48 & 0.22 & -0.04 \\
    27 & \ce{C3H3}  & -0.98 & -0.88 & -0.73 & -0.10 & -2.29 & -1.61 & -0.87 & -0.68 & 1.31 & 0.73 & 0.14  \\
    31 & \ce{C2H3F} & -2.21 & -2.09 & -1.92 & -0.12 & -2.50 & -2.22 & -1.91 & -0.28 & 0.29 & 0.13 & -0.01 \\
    39 & \ce{SiH4}  & -0.92 & -0.83 & -0.72 & -0.09 & -1.75 & -1.42 & -1.08 & -0.33 & 0.83 & 0.59 & 0.36  \\
    47 & \ce{NH3}   & -0.93 & -0.85 & -0.76 & -0.08 & -1.85 & -1.35 & -0.85 & -0.51 & 0.93 & 0.50 & 0.09  \\
    66 & \ce{HCN}   & -2.59 & -2.48 & -2.31 & -0.12 & -2.96 & -2.61 & -2.23 & -0.35 & 0.37 & 0.14 & -0.08 \\
    70 & \ce{CH4O}  & -1.05 & -0.95 & -0.81 & -0.10 & -2.00 & -1.49 & -0.93 & -0.52 & 0.95 & 0.54 & 0.12  \\
    71 & \ce{C2H6O} & -0.97 & -0.87 & -0.73 & -0.10 & -1.90 & -1.39 & -0.84 & -0.51 & 0.93 & 0.52 & 0.11  \\
    76 & \ce{H2O}   & -1.02 & -0.96 & -0.88 & -0.06 & -1.75 & -1.34 & -0.89 & -0.41 & 0.73 & 0.38 & 0.01  \\
             \hline
            &  MD &        &        &        &  0.09  &       &      &       & 0.48  & 0.81 & 0.43 & 0.05  \\
        \bottomrule
    \end{tabular}
    \caption{Comparison of EAs for selected molecules from the GW100 database calculated with ADF with and without diffuse functions. All values are in eV.}
    \label{tab::diffuseVsnot-diffuse}
\end{table}

One furthermore clearly sees that for a group of molecules with significant deviations from ADF to def2-GTO of up to several eVs. These are the already mentioned systems with positive LUMO. Contrariwise, the raw data in table~\ref{tab::EAs} shows that the agreement with WEST is especially good for systems with positive LUMO. As an example, consider the series of linear alkane chains, $\text{C}_n\text{H}_{2n+1}$ for $n = 1, \dots, 4$. With deviations from 10 to 40 meV, the agreement with WEST is excellent, while def2-GTO overestimates the EAs of these systems by more than 1 eV. In this context, it is interesting to investigate the effect of the diffuse functions. This is shown in table~\ref{tab::diffuseVsnot-diffuse} for some systems with LUMO well above the vacuum level. Only comparing the basis set extrapolated values, the effect of the diffuse functions seem to be rather small; for \ce{N2} the EA calculated from the basis set without the diffuse functions is with 250 meV even higher than the augmented basis sets, and overall, the average difference is only 50 meV which is well within the expected error range from the CBS limit extrapolation. However, comparing the results form the finite basis sets, the differences are exorbitant. Especially, on the TZ level, the addition of diffuse functions results in a lowering of the EAs by nearly 1 eV on average. For the non-augmented basis sets, the average difference between TZ and QZ basis set is 480 meV, resulting in differences of sometimes more than 1 eV between the EAs on the TZ and on the extrapolated level. In light of these differences the good agreement between the CBS limit extrapolated EAs is remarkable. Despite this good agreement, the augmented basis sets should be the preferred choice to calculate EAs of systems with unbound LUMOs since the extrapolation procedure is generally less reliable when the differences between the results for the finite basis sets are larger.

\begin{figure}[ht]
    \centering
    \includegraphics[width=0.8\textwidth]{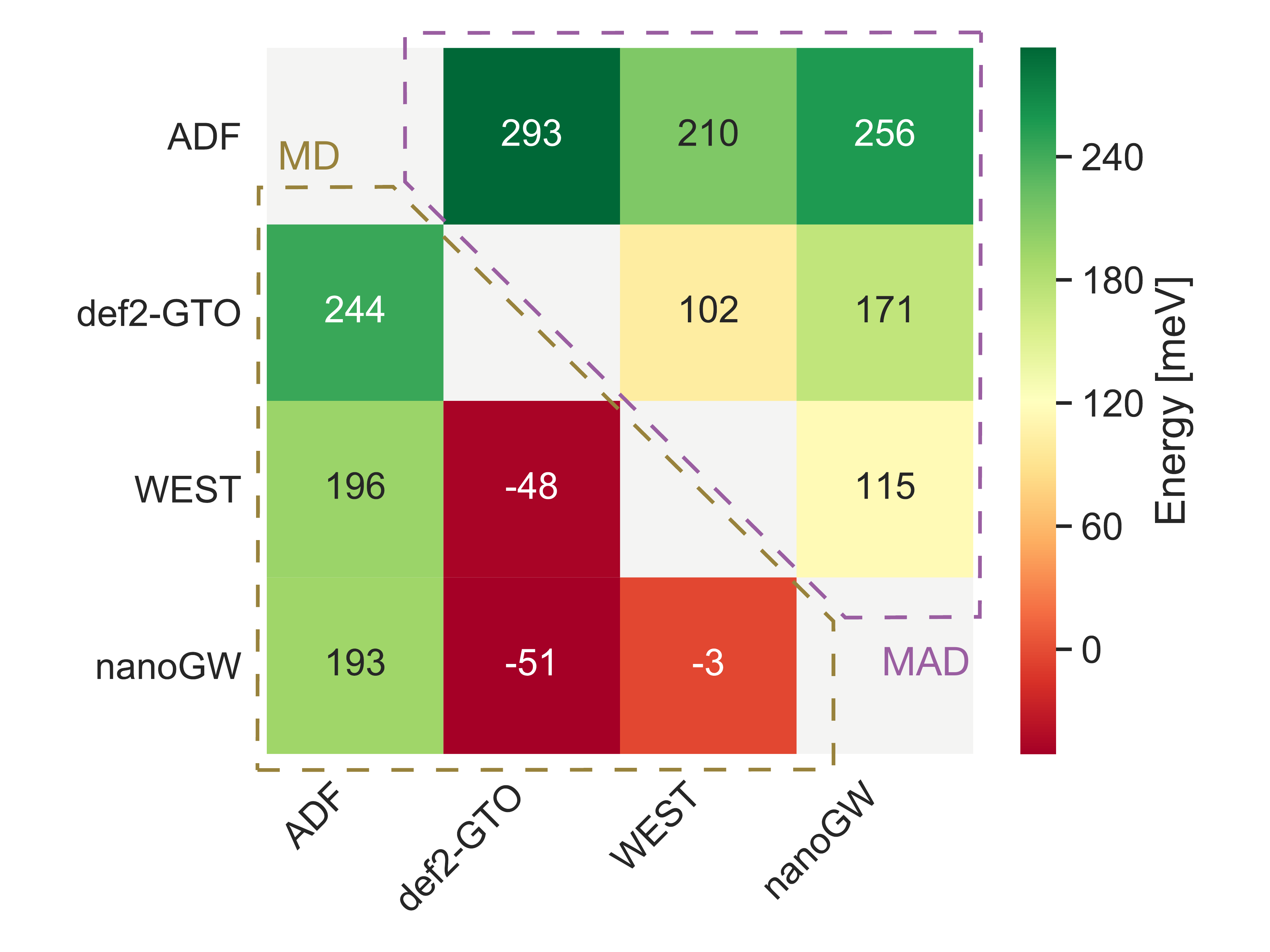}
    \caption{MADs, (upper triangle) and mean deviations (MD) (lower triangle) of the CBS limit extrapolated EAs of the subset of systems with bound LUMO and medium sized organic molecules (in total 48 data points) in the GW100 dataset computed with different codes. All values are in eV.}
    \label{fig::MD_LUMO}
\end{figure}

Finally, we compare our EAs for systems with a bound LUMO to WEST, nanoGW, and def-GTO results. The MADs and MDs in figure~\ref{fig::MD_LUMO} show that def2-GTO, nanoGW and and WEST are generally in good agreement for these systems while the MADs to ADF are large. As the MDs show, ADF significantly overestimates these EAs compared to the other codes, which indicates that the results are not entirely converged to the CBS limit. This interpretation is also in line with the raw data in table~\ref{tab::EAs} showing that the deviations are generally largest for di- and triatomic systems as well as molecules containing Fluorine and Chlorine, while the agreement for the medium organic molecules like the nucleo-bases is satisfactory. Adding additional diffuse or tight functions would possible not result in an improved description of the EAs of the former systems. Instead, reaching the CBS limit is most likely only possible using basis functions with higher angular momenta than $l=3$ which are not available to us. However, we can try to simulate the effect of these functions by adding additional off-center Slater functions to the basis set which can be achieved conveniently by adding ghost atoms. This approach is reminiscent of bond-centred basis functions which have for example been used by Dunlap and coworkers.\cite{Dunlap1983a}

\begin{table}
    \centering
    \begin{tabular}{lcccccc}
       \toprule 
        system  & T & Q & ex. & ex. (no gh.) & def2-GTO & WEST \\
       \midrule
        \ce{F2}   & 0.20 & 0.51 & 0.90 & 0.54 & 1.23 & 1.06  \\
        \ce{Cl2}  & 0.74 & 0.95 & 1.21 & 0.83 & 1.40 & 1.38  \\
        \ce{Br2}  & 1.27 & 1.45 & 1.69 & 1.40 & 1.96 & 1.88  \\
        \ce{TiF4} &-0.33 &-0.10 & 0.19 & 0.09 & 1.06 & 0.92  \\
        \ce{SO2}  & 0.84 & 1.07 & 1.38 & 0.86 & 1.49 & 1.37  \\
        \bottomrule
    \end{tabular}
    \caption{Effect of the addition of additional off-center Slater functions via Ne ghost atoms on the EAs of selected systems form GW100. All values are in eV.}
    \label{tab::ghost}
\end{table}

As examples, we consider the dihalogens \ce{F2}, \ce{Cl2} and \ce{Br2}, \ce{SO2}, as well as \ce{TiF4} and we augment these structures with Ne ghost atoms for which we use the same basis set than for the real atoms. For each atom $A$ in a systems, we place two ghost atoms $G, G'$ on a straight line defined by the position of $A$ and every neighboring atom $B$ so that the distance between $A$ and $G$ ($G'$) equals one third (minus one third) of the distance between $A$ and $B$. The results of this augmentation is shown in table~\ref{tab::ghost} and we clearly see that it reduces the basis set error considerably, as compared to the other codes. Of course, such an augmentation should not be seen as a practical solution but it shows that agreement between ADF and the other codes can in principle be reached also for these systems.

\subsection{Ionization Potentials and Electron Affinities for a Subset of GW5000}

For the GW100 database, discrepancies to other codes are more pronounced for smaller systems while the agreement for medium systems is significantly better. To confirm this observation for a larger number of systems, we also calculated IPs and EAs of a subset of 250  medium to large organic molecules from the GW5000 database\cite{Stuke2020} for which CBS limit extrapolated references values calculated with FHI-AIMS using the def2-GTO basis sets are available. We have already considered this subset in ref~\cite{Forster2020b}. We used here the TZ3P and QZ6P basis sets without diffuse functions since the LUMOs of the considered systems are negative. As expected, we find better agreement with FHI-AIMS than for GW100: For IPs, ADF deviated to FHI-AIMS by 62 meV on average as opposed to the 86 meV for GW100. For EAs, the MAD is with 93 meV slightly worse but the agreement is much better than for GW100.

\section{\label{sec::conclusion}Conclusions}

The GW100 benchmark by van Setten et al.\cite{VanSetten2015} has become the most important resource for comparing different implementations of the $GW$ method for finite systems. IPs and EAs calculated with implementations using def2 GTO-type basis sets,\cite{VanSetten2015} plane waves,\cite{Maggio2017a, Govoni2018} and real-space finite elements\cite{Gao2019} have been compared in the recent past and good agreement between CBS limit extrapolated results has been found for IPs and EAs of molecules with bound LUMO. In this work, we extended the list of available results and calculated the IPs and EAs in the GW100 dataset using Slater type orbitals. For this purpose we have developed new Slater type basis sets which allow us to extrapolate our results to the CBS limit. These basis sets are available online.\bibnote{The basis set files can be downloaded from \href{http://www.scm.com/wp-content/uploads/Correlation-basis-sets.zip}{http://www.scm.com/wp-content/uploads/Correlation-basis-sets.zip}}

Our study confirms once more that it is possible, though difficult, to reach consensus between different implementations. All implementations compared in this work do not only implement the GWA with different basis sets, but they also differ in other technical parameters like frequency treatment, description of core electrons, the algorithm used to solve the QP equations and the numerical treatment of 4-point correlation functions. In light of these differences, the observed agreement between 55 to 85 meV on average for IPs between STO- def2-GTO- and plane-wave results is excellent. Reaching the CBS limit is more difficult for EAs than for IPs. Still, EAs calculated with ADF are in excellent agreement with the place-wave results from the WEST code for systems with positive LUMOs, with an overall MAD between both codes of 160 meV. These deviations mostly stem from large basis set errors for the EAs of small molecules with bound LUMO with the ADF code. For larger, organic molecules, agreement to def2-GTO results is with a MAD of 93 meV significantly better.

We observe that ADF tends to give lower EAs and IPs than the def2-GTO and plane-wave results. This is an indicator that the results are not entirely converged with respect to the CBS limit. Our STO-type basis sets are restricted to angular momenta smaller or equal to $l=3$ which sets a limit to the accuracy currently attainable with STOs. It would be desirable to obtain results for the GW100 database using even larger STO type basis sets containing also functions with higher angular momenta. This would enable a more accurate estimate of the CBS limit and might hopefully lead to an even better agreement with other implementations. On the other hand, the best IPs for GW100 in our earlier work showed a MAD of more than 300 meV to def-GTO and nanoGW.\cite{Forster2020b} Thus, the improvements which are due to our improved basis sets and imaginary time and frequency grids are already immense. 

The close agreement between different codes is highly important in practice since it allows researchers to interpret the results of their $GW$ calculations, without worrying that they might be skewed by technical aspects. The slow convergence of $GW$ calculations to the CBS limit remains an obstacle in practical applications. Especially for $GW$ calculations for finite systems with hundreds of atoms, which have become the focus of much research in the last years\cite{Wilhelm2018,Wilhelm2021,Forster2020b,Duchemin2021}, calculations at (or even close to) the CBS limit for individual QP energies are currently often out of reach and will most likely not become routine anytime soon. To overcome this issue, finite basis set corrections\cite{Bruneval2016, Riemelmoser2020, Loos2020a, Bruneval2020} hold much promise and it is to hope that these techniques will be further developed and become more widespread available in the near future. For differences of QP energies, the situation is already much better. When augmented basis sets are used, basis set errors usually cancel to a large extent and already on the DZ level fundamental gaps are often sufficiently converged to the CBS limit.\cite{Blase2011, Boulanger2014,
Faber2015,
Jacquemin2015,
Bruneval2015,
Wilhelm2016a,
Forster2020b}


\appendix

\section{\label{app::A}Implementation of Imaginary Time and Frequency Grids}

In this appendix we state the formulas we use to calculate imaginary time and imaginary frequency grids. For derivations, we refer to ref.~\citen{Kaltak2014}. Since we want to switch between imaginary time and imaginary frequency, we write the discrete Fourier transform $f(i\omega_k)$ of an arbitrary function $f(i\tau_j)$ as
\begin{equation}
 \label{exactFFT}
 f(i\omega_k) =
 -i\sum^{N_{\tau}}_{j}\left\{\gamma^{(c)}_{kj}\cos (\omega_k \tau_j) 
 \left(f(i \tau_j) + f(-i \tau_j)\right) - 
 i \gamma^{(s)}_{kj}\sin (\omega_k \tau_j) 
 \left(f(i \tau_j) - f(-i \tau_j)\right)\right\} \;.
 \end{equation}
Fourier transform from imaginary frequency to imaginary time can be done by inverting the matrices with the elements $\gamma^{(c)}_{kj}\cos (\omega_k \tau_j)$ and $\gamma^{(s)}_{kj}\sin (\omega_k \tau_j)$ (In case $N_{\tau} \neq N_{\omega}$, a pseudo-inverse can be calculated). At the beginning of a GW calculation, we find optimal sets of imaginary frequency $\left\{\omega_k)\right\}_{k = 1, \dots N_{\omega}}$ and imaginary time points $\left\{\tau_j)\right\}_{j = 1, \dots N_{\tau}}$. For imaginary time, we use the algorithm described in ref.~\citen{Helmich-Paris2016} to minimize 
\begin{equation}
    \left\Vert \eta^{(\tau)}(x;\alpha,\tau) \right\Vert_{\infty} \;,
    \quad \eta^{(\tau)}(x;\alpha,\tau) = 
    \frac{1}{2x} - \sum^{N_{\tau}}_j \alpha_j e^{-2x\tau_j}. \;,
\end{equation}
and for imaginary frequency, we first minimize 
\begin{equation}
\label{2}
    \left\Vert\eta^{(\omega)}(x;\sigma,\omega)\right\Vert_{2} \;, \quad
    \eta^{(\omega)}(x;\sigma,\omega) = 
    \frac{1}{x} - \frac{1}{\pi} \sum^{N_{\omega}}_k \sigma_k \left(\frac{2x}{x^2 + \omega^2_k}\right)^2 \;,
\end{equation}
using a Levenberg–Marquardt algorithm (LMA). In both minimizations, $x$ denotes the electron-hole pair transition energy Both, $\eta^{(\tau)}$ and $\eta^{(\omega)}$ are minimized in the interval $[1,x_{min}/x_{max}]$, with $x_{min}$ ($x_{max}$) being the smallest (largest) considered electron-hole transition energy. In both optimizations, the interval $[x_{min},x_{max}]$ is represented on a logarithmic grid. After minimizing the $L_2$ norm on the logarithmic grid, we minimize the $L_2$ norm for the positions of the minima and maximima and repeat this procedure until self-consistency is reached. The coefficients $\left\{\sigma_k)\right\}_{k = 1, \dots N_{\omega}}$ and $\left\{\alpha_j)\right\}_{j = 1, \dots N_{\tau}}$ are byproducts of this optimization but they can be used to evaluate MP2 or RPA correlation energies. 

$\eta^{(\omega)}$ has multiple minima and especially for larger $N_{\omega}$ the LMA only converges to a useful minimum when it is initialized with $\sigma$ and $\omega$ which are already sufficiently close to the parameters which minimize $\eta^{(\omega)}$. In practice, we found that at a useful (but not necessarily global) minimum, $\eta^{(\omega)}$ has $2N_{\omega}$ or $2N_{\omega} -1$ extrema. This behaviour can be exploited to find good starting points for the LMA for different ratios $x_{min}/x_{max}$ and $N_{\omega}$. We always start the LMA from pretabulated values.\bibnote{The pretabulated values are available on Github (\href{https://github.com/ArnoFoerster/Imaginary-Frequency-Grids-GW-and-RPA}{https://github.com/ArnoFoerster/Imaginary-Frequency-Grids-GW-and-RPA})} These values have been obtained by a simple metropolis algorithm for several $x_{min}/x_{max}$ and for $N_{\omega}$ between 1 and 40 and are chosen so that $\eta^{(\omega)}$ has $2N_{\omega}$ or $2N_{\omega} -1$ extrema for a given  $N_{\omega}$. In a GWA calculation, the LMA is then initialized with the pretabulated parameters which are closest to the $x_{min}/x_{max}$ of the calculation. 

To avoid unnecessarily large grids, the $N_{\omega}$ and $N_{\tau}$ are determined at run-time and the user only specifies an upper limit of points for both, imaginary time and frequency grids. More precisely, for small $N_{\omega}$, grid points and weights are calculated and the $L_2$ norm of $\eta^{\omega}$ is calculated. Then we increase the imaginary frequency grid until the least square error is smaller than $\epsilon_{\omega} = 1e^{-10}$. The number of points which are required to reach that accuracy strongly depends on $x_{min}/x_{max}$. In our experience, the QP energies converge faster with respect to the imaginary time grid than with respect to the $\omega$-grid, Since we find it convenient to work with grids of the same size (although this is not necessary), we set $N_{\tau} = N_{\omega}$. For example, for a Hydrogen-molecule in a triple-$\zeta$ (TZ) basis, $N_{\omega} = 7$ will already be sufficient to reach the desired accuracy, while for the Iodine molecule in a QZ basis $N_{\omega} = 31$ will be necessary.

To calculate the matrices $\gamma^{(c)}$ we minimize
\begin{equation}
    \left\Vert\eta^{(c)}\left(x;\gamma^{(c)}\right)\right\Vert_2 \;, \quad 
    \eta^{(c)}\left(x,\gamma^{(c)}\right) = \frac{2x}{x^2 + \omega^2_k} - \sum^{N_{\tau}}_{j=1} 
    \gamma^{(c)}_{kj}\cos (\omega_k \tau_j)e^{-x\tau_j} \;,
\end{equation}
for all $\omega_k$ and for $\gamma^{(s)}$,  we minimize
\begin{equation}
    \left\Vert\eta^{(s)}\left(x;\gamma^{(s)}\right)\right\Vert_2 \;, \quad 
    \eta^{(s)}\left(x,\gamma^{(s)}\right) = \frac{2\omega_k}{x^2 + \omega^2_k} - \sum^{N_{\tau}}_{j=1} 
    \gamma^{(c)}_{kj}\sin (\omega_k \tau_j)e^{-x\tau_j} \;,
\end{equation}
with a LMA. The procedure is the same as described in ref.~\citen{Liu2016}. In all cases, the algorithm converges smoothly from arbitrary starting values. 

\begin{acknowledgement}
This research received funding from the Netherlands Organisation for Scientific Research (NWO) in the framework of the Innovation Fund for Chemistry and from the Ministry of Economic Affairs in the framework of the \enquote{\emph{TKI/PPS-Toeslagregeling}}. The authors thank Erik van Lenthe for fruitful discussions.
\end{acknowledgement}


\begin{suppinfo}
KS HOMO and LUMO eigenvalues, QP energies for finite basis sets, raw data for the subset of 250 molecules from GW5000 and technical parameters of all calculations for GW100. 
\end{suppinfo}


\bibliography{all}


\begin{tocentry}
\includegraphics[width=\textwidth]{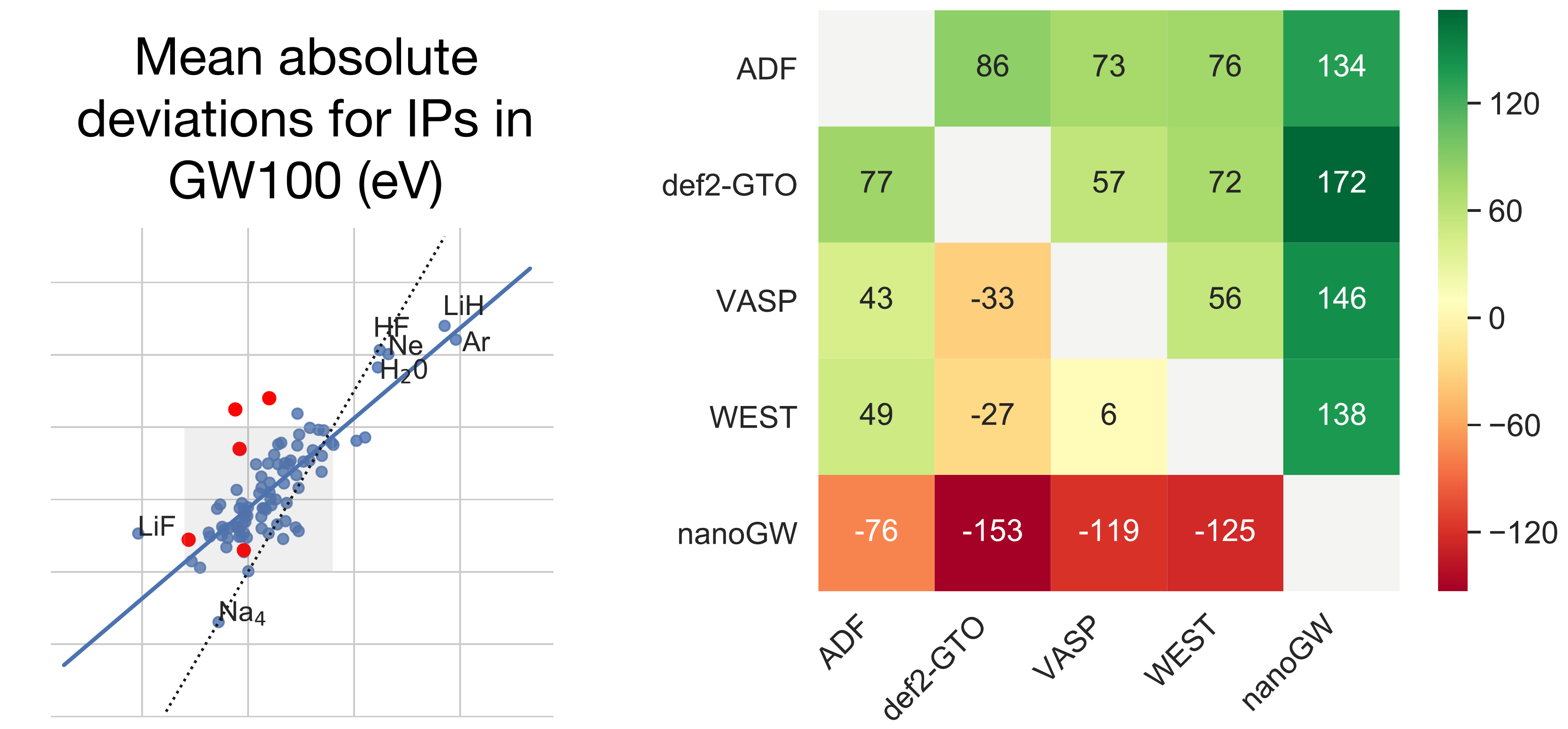}
\end{tocentry}



\section*{Supplementary material}

# Supporting information for: GW100: A Slater Type Orbital Perspective with ADF

Arno Förster* and Lucas Visscher

*Theoretical Chemistry, Vrije Universiteit, De Boelelaan 1083, NL-1081 HV, Amsterdam, The Netherlands*

E-mail: a.t.l.foerster@vu.nl

The following three tables contain additional information on the calculation we have performed. The first table contains the KS eigenvalues calculated with PBE, and the second and third table contain the quasi-particle energies at the aug-TZ3P and aug-QZ6P level of theory, respectively. They also list technical parameters which we have used in the calculations.

Table 1: PBE HOMO and LUMO energies at finite basis sets (all values in eV). TZ denotes aug-TZ3P and QZ denotes aug-QZ6P

| | | HOMO | | LUMO | |
|---|---|---|---|---|---|
| | Name | TZ | QZ | TZ | QZ |
| 1 | Helium | −15.76 | −15.76 | 1.41 | 0.34 |
| 2 | Neon | −13.36 | −13.35 | 0.59 | −0.02 |
| 3 | Argon | −10.29 | −10.29 | 0.29 | −0.28 |
| 4 | Krypton | −9.30 | −9.28 | −0.16 | −0.40 |
| 5 | Xenon | −8.30 | −8.29 | −0.30 | −0.43 |
| 6 | Hydrogen | −10.39 | −10.38 | 0.29 | 0.26 |
| 7 | Lithiumdimer | −3.23 | −3.22 | −1.79 | −1.79 |
| 8 | Sodiumdimer | −3.13 | −3.13 | −1.78 | −1.78 |
| 9 | Sodiumtetramer | −2.68 | −2.68 | −2.09 | −2.09 |
| 10 | Sodiumhexamer | −2.99 | −2.99 | −1.89 | −1.89 |
| 11 | Dipotassium | −2.56 | −2.56 | −1.61 | −1.61 |
| 12 | Dirubidium | −2.44 | −2.44 | −1.54 | −1.54 |
| 13 | Nitrogen | −10.27 | −10.27 | −1.96 | −1.96 |
| 14 | Phosphorusdimer | −7.14 | −7.13 | −3.43 | −3.43 |
| 15 | Arsenicdimer | −6.59 | −6.57 | −3.45 | −3.44 |
| 16 | Fluorine | −9.46 | −9.45 | −5.82 | −5.80 |
| 17 | Chlorine | −7.30 | −7.31 | −4.24 | −4.24 |

Continued on next page

1

|  | Name | HOMO | | LUMO | |
|---|---|---|---|---|---|
|  |  | TZ | QZ | TZ | QZ |
| 18 | Bromine | $-6.86$ | $-6.85$ | $-4.48$ | $-4.46$ |
| 19 | Iodine | $-6.34$ | $-6.33$ | $-4.33$ | $-4.32$ |
| 20 | Methane | $-9.46$ | $-9.46$ | $-0.39$ | $-0.39$ |
| 21 | Ethane | $-8.16$ | $-8.16$ | $-0.44$ | $-0.45$ |
| 22 | Propane | $-7.76$ | $-7.76$ | $-0.48$ | $-0.49$ |
| 23 | Buthane | $-7.58$ | $-7.58$ | $-0.49$ | $-0.50$ |
| 24 | Ethylene | $-6.78$ | $-6.78$ | $-1.07$ | $-1.07$ |
| 25 | Acetylene | $-7.20$ | $-7.20$ | $-0.42$ | $-0.42$ |
| 26 | Tetracarbon | $-7.27$ | $-7.26$ | $-6.12$ | $-6.11$ |
| 27 | Cyclopropane | $-7.05$ | $-7.05$ | $-0.34$ | $-0.36$ |
| 28 | Benzene | $-6.34$ | $-6.34$ | $-1.26$ | $-1.25$ |
| 29 | Cyclooctatetraene | $-5.31$ | $-5.30$ | $-2.32$ | $-2.32$ |
| 30 | Cyclopentadiene | $-5.41$ | $-5.41$ | $-1.49$ | $-1.49$ |
| 31 | Vynilfluoride | $-6.55$ | $-6.55$ | $-0.97$ | $-0.97$ |
| 32 | Vynilchloride | $-6.44$ | $-6.44$ | $-1.43$ | $-1.44$ |
| 33 | Vynilbromide | $-5.86$ | $-5.85$ | $-1.37$ | $-1.37$ |
| 34 | Vyniliodide | $-6.10$ | $-6.09$ | $-1.71$ | $-1.70$ |
| 35 | Carbontetrafluoride | $-10.43$ | $-10.41$ | $-0.42$ | $-0.43$ |
| 36 | Carbontetrachloride | $-7.67$ | $-7.68$ | $-2.78$ | $-2.77$ |
| 37 | Carbontetrabromide | $-7.00$ | $-6.99$ | $-3.56$ | $-3.53$ |
| 38 | Carbontetraiodide | $-6.29$ | $-6.28$ | $-4.29$ | $-4.28$ |
| 39 | Silane | $-8.53$ | $-8.52$ | $-0.47$ | $-0.49$ |
| 40 | Germane | $-8.38$ | $-8.37$ | $-0.67$ | $-0.68$ |
| 41 | Disilane | $-7.30$ | $-7.29$ | $-0.68$ | $-0.69$ |
| 42 | Pentasilane | $-6.59$ | $-6.58$ | $-1.68$ | $-1.67$ |
| 43 | Lithiumhydride | $-4.36$ | $-4.36$ | $-1.62$ | $-1.64$ |
| 44 | Potassiumhydride | $-3.46$ | $-3.46$ | $-1.62$ | $-1.62$ |
| 45 | Borane | $-8.50$ | $-8.49$ | $-3.07$ | $-3.07$ |
| 46 | Diborane6 | $-7.87$ | $-7.87$ | $-2.04$ | $-2.04$ |
| 47 | Amonia | $-6.19$ | $-6.18$ | $-0.74$ | $-0.75$ |
| 48 | Hydrogenazide | $-6.81$ | $-6.80$ | $-2.10$ | $-2.10$ |
| 49 | Phosphine | $-6.72$ | $-6.72$ | $-0.67$ | $-0.67$ |
| 50 | Arsine | $-6.73$ | $-6.74$ | $-0.77$ | $-0.77$ |
| 51 | Hydrogensulfide | $-6.30$ | $-6.30$ | $-0.86$ | $-0.87$ |
| 52 | Hydrogenfluoride | $-9.66$ | $-9.65$ | $-0.97$ | $-0.97$ |
| 53 | Hydrogenchloride | $-8.04$ | $-8.04$ | $-1.12$ | $-1.12$ |
| 54 | Lithiumfluoride | $-6.13$ | $-6.13$ | $-1.53$ | $-1.52$ |
| 55 | Magnesiumfluoride | $-8.30$ | $-8.30$ | $-2.58$ | $-2.58$ |
| 56 | Titaniumfluoride | $-10.45$ | $-10.44$ | $-4.19$ | $-4.19$ |
| 57 | Aluminumtrifluoride | $-9.72$ | $-9.71$ | $-2.54$ | $-2.54$ |
| 58 | Fluoroborane | $-6.79$ | $-6.78$ | $-2.15$ | $-2.14$ |
| 59 | Sulfertetrafluoride | $-8.25$ | $-8.24$ | $-2.97$ | $-2.96$ |
| 60 | Potassiumbromide | $-4.76$ | $-4.76$ | $-1.87$ | $-1.87$ |
| 61 | Galliummonochloride | $-6.53$ | $-6.53$ | $-2.44$ | $-2.43$ |
| 62 | Sodiumchloride | $-5.29$ | $-5.29$ | $-2.24$ | $-2.24$ |
| 63 | Magnesiumchloride | $-7.63$ | $-7.63$ | $-2.55$ | $-2.54$ |
| 64 | Aluminumtriiodide | $-6.72$ | $-6.71$ | $-2.71$ | $-2.70$ |
| 65 | Boronnitride | $-7.46$ | $-7.46$ | $-7.29$ | $-7.29$ |
| 66 | Hydrogencyanide | $-9.04$ | $-9.04$ | $-1.11$ | $-1.11$ |
| 67 | Phosphorusmononitride | $-7.77$ | $-7.76$ | $-3.41$ | $-3.40$ |
| 68 | Hydrazene | $-5.30$ | $-5.30$ | $-0.96$ | $-0.96$ |
| 69 | Formaldehyde | $-6.28$ | $-6.27$ | $-2.71$ | $-2.71$ |
| 70 | Methanol | $-6.35$ | $-6.35$ | $-0.65$ | $-0.66$ |

|      |                      | HOMO |       | LUMO |       |
|------|----------------------|------|-------|------|-------|
|      | Name                 | TZ   | QZ    | TZ   | QZ    |
| 71   | Ethanol              | −6.16 | −6.16 | −0.67 | −0.68 |
| 72   | Acetaldehyde         | −5.98 | −5.98 | −2.16 | −2.16 |
| 73   | Ethoxyethane         | −5.81 | −5.80 | −0.51 | −0.53 |
| 74   | FormicAcid           | −6.95 | −6.94 | −1.56 | −1.56 |
| 75   | Hydrogenperoxide     | −6.46 | −6.45 | −1.69 | −1.68 |
| 76   | Water                | −7.26 | −7.25 | −0.94 | −0.93 |
| 77   | Carbondioxide        | −9.10 | −9.09 | −0.91 | −0.95 |
| 78   | Carbondisulfide      | −6.81 | −6.80 | −2.86 | −2.86 |
| 79   | Carbonoxysulfide     | −7.49 | −7.48 | −1.95 | −1.95 |
| 80   | Carbonoxyselenide    | −6.99 | −6.98 | −2.07 | −2.07 |
| 81   | Carbonmonoxide       | −9.35 | −9.34 | −3.35 | −3.35 |
| 82   | Ozon                 | −7.96 | −7.96 | −6.16 | −6.16 |
| 83   | Sulferdioxide        | −8.09 | −8.08 | −4.41 | −4.40 |
| 84   | Berylliummonoxide    | −6.14 | −6.14 | −4.81 | −4.81 |
| 85   | Magnesiummonoxide    | −4.80 | −4.80 | −4.29 | −4.29 |
| 86   | Tuloene              | −6.01 | −6.01 | −1.23 | −1.23 |
| 87   | Ethybenzene          | −6.01 | −6.01 | −1.16 | −1.16 |
| 88   | Hexafluorobenzene    | −6.66 | −6.66 | −2.22 | −2.22 |
| 89   | Phenol               | −5.64 | −5.64 | −1.37 | −1.36 |
| 90   | Aniline              | −5.03 | −5.03 | −1.12 | −1.12 |
| 91   | Pyridine             | −5.96 | −5.95 | −1.91 | −1.90 |
| 92   | Guanine              | −5.30 | −5.29 | −1.43 | −1.43 |
| 93   | Adenine              | −5.53 | −5.53 | −1.71 | −1.71 |
| 94   | Cytosine             | −5.73 | −5.73 | −2.07 | −2.07 |
| 95   | Thymine              | −6.06 | −6.05 | −2.29 | −2.28 |
| 96   | Uracil               | −6.29 | −6.28 | −2.45 | −2.44 |
| 97   | Urea                 | −5.94 | −5.93 | −1.01 | −1.02 |
| 98   | Silverdimer          | −4.75 | −4.77 | −2.75 | −2.76 |
| 98   | Silverdimer (ZORA)   | −5.21 | −3.11 |      |       |
| 99   | Copperdimer          | −4.78 | −4.78 | −2.94 | −3.00 |
| 100  | Coppercyanide        | −6.69 | −6.72 | −4.00 | −4.04 |

Table 2: IPs and EAs on the aug-TZ3P level of theory (in eV) and technical parameters used in the calculations: Number of grid points, number of orbitals and fit set (N = *Normal*, G = *Good*, VG = *VeryGood*).

|      | Name           | IP    | EA    | $N_\omega$ | $N_\tau$ | $N_{bas}$ | fit set |
|------|----------------|-------|-------|------------|----------|-----------|---------|
| 1    | Helium         | 23.26 | −2.89 | 10         | 10       | 18        | N       |
| 2    | Neon           | 20.09 | −1.77 | 15         | 15       | 40        | N       |
| 3    | Argon          | 14.49 | −2.27 | 18         | 18       | 48        | N       |
| 4    | Krypton        | 13.20 | −1.52 | 22         | 22       | 70        | N       |
| 5    | Xenon          | 11.70 | −1.18 | 27         | 31       | 100       | N       |
| 6    | Hydrogen       | 15.59 | −0.90 | 7          | 7        | 36        | N       |
| 7    | Lithiumdimer   | 4.78  | 0.36  | 17         | 17       | 74        | N       |
| 8    | Sodiumdimer    | 4.74  | 0.41  | 20         | 20       | 90        | N       |
| 9    | Sodiumtetramer | 4.15  | 0.92  | 22         | 22       | 180       | N       |
| 10   | Sodiumhexamer  | 4.20  | 0.95  | 21         | 21       | 270       | N       |
| 11   | Dipotassium    | 3.89  | 0.48  | 24         | 24       | 116       | N       |
| 12   | Dirubidium     | 3.75  | 0.53  | 27         | 30       | 168       | N       |
| 13   | Nitrogen       | 14.66 | −2.65 | 16         | 16       | 80        | N       |

|    | Name | IP | EA | $N_\omega$ | $N_\tau$ | $N_{bas}$ | fit set |
|----|------|-----|-----|-----|-----|-----|-----|
| 14 | Phosphorusdimer | 9.88 | 0.38 | 20 | 20 | 96 | N |
| 15 | Arsenicdimer | 9.07 | 0.50 | 24 | 24 | 140 | N |
| 16 | Fluorine | 14.62 | −0.14 | 17 | 17 | 80 | N |
| 17 | Chlorine | 10.73 | 0.33 | 21 | 21 | 96 | N |
| 18 | Bromine | 9.91 | 0.90 | 24 | 24 | 140 | N |
| 19 | Iodine | 9.01 | 1.28 | 27 | 29 | 200 | N |
| 20 | Methane | 13.80 | −0.97 | 14 | 14 | 112 | VG |
| 21 | Ethane | 12.23 | −0.96 | 15 | 15 | 188 | VG |
| 22 | Propane | 11.64 | −0.92 | 15 | 15 | 264 | VG |
| 23 | Buthane | 11.37 | −0.89 | 15 | 15 | 340 | VG |
| 24 | Ethylene | 10.09 | −2.12 | 16 | 16 | 152 | VG |
| 25 | Acetylene | 10.83 | −2.76 | 16 | 16 | 116 | VG |
| 26 | Tetracarbon | 10.56 | 2.27 | 19 | 19 | 160 | G |
| 27 | Cyclopropane | 10.39 | −0.98 | 16 | 16 | 228 | VG |
| 28 | Benzene | 8.82 | −1.26 | 17 | 17 | 348 | G |
| 29 | Cyclooctatetraene | 7.92 | −0.30 | 18 | 18 | 464 | G |
| 30 | Cyclopentadiene | 8.13 | −1.17 | 18 | 18 | 308 | G |
| 31 | Vynilfluoride | 9.98 | −2.21 | 18 | 18 | 174 | G |
| 32 | Vynilchloride | 9.48 | −1.61 | 20 | 20 | 182 | G |
| 33 | Vynilbromide | 8.76 | −1.50 | 23 | 23 | 204 | N |
| 34 | Vyniliodide | 8.82 | −1.06 | 27 | 32 | 234 | N |
| 35 | Carbontetrafluoride | 15.10 | −0.92 | 16 | 16 | 200 | G |
| 36 | Carbontetrachloride | 10.68 | −0.41 | 20 | 20 | 232 | G |
| 37 | Carbontetrabromide | 9.64 | 0.62 | 24 | 24 | 320 | N |
| 38 | Carbontetraiodide | 8.63 | 1.64 | 27 | 31 | 440 | N |
| 39 | Silane | 12.17 | −0.92 | 17 | 17 | 120 | VG |
| 40 | Germane | 11.85 | −1.00 | 22 | 22 | 143 | G |
| 41 | Disilane | 10.11 | −0.93 | 18 | 18 | 204 | G |
| 42 | Pentasilane | 8.73 | −0.39 | 19 | 19 | 456 | G |
| 43 | Lithiumhydride | 5.98 | −0.02 | 16 | 16 | 55 | N |
| 44 | Potassiumhydride | 4.83 | 0.06 | 23 | 23 | 76 | N |
| 45 | Borane | 12.68 | −0.48 | 15 | 15 | 94 | G |
| 46 | Diborane6 | 11.75 | −1.07 | 16 | 16 | 188 | G |
| 47 | Amonia | 10.00 | −0.93 | 17 | 17 | 94 | G |
| 48 | Hydrogenazide | 10.16 | −1.67 | 16 | 16 | 138 | G |
| 49 | Phosphine | 9.99 | −0.84 | 19 | 19 | 102 | G |
| 50 | Arsine | 9.87 | −0.84 | 22 | 22 | 124 | G |
| 51 | Hydrogensulfide | 9.73 | −0.94 | 19 | 19 | 84 | N |
| 52 | Hydrogenfluoride | 14.88 | −1.22 | 16 | 16 | 58 | N |
| 53 | Hydrogenchloride | 11.81 | −1.38 | 19 | 19 | 66 | N |
| 54 | Lithiumfluoride | 9.80 | −0.08 | 15 | 15 | 77 | N |
| 55 | Magnesiumfluoride | 12.16 | 0.08 | 17 | 17 | 128 | N |
| 56 | Titaniumfluoride | 13.67 | −0.89 | 20 | 20 | 223 | N |
| 57 | Aluminumtrifluoride | 14.04 | −0.21 | 18 | 18 | 168 | G |
| 58 | Fluoroborane | 10.21 | −1.49 | 16 | 16 | 80 | N |
| 59 | Sulfertetrafluoride | 11.78 | −0.73 | 19 | 19 | 208 | G |
| 60 | Potassiumbromide | 7.24 | 0.23 | 24 | 24 | 128 | N |
| 61 | Galliummonochloride | 9.28 | −0.24 | 24 | 24 | 119 | N |
| 62 | Sodiumchloride | 8.02 | 0.30 | 21 | 21 | 93 | N |
| 63 | Magnesiumchloride | 10.67 | 0.27 | 19 | 19 | 144 | N |
| 64 | Aluminumtriiodide | 9.17 | 0.39 | 27 | 30 | 348 | N |
| 65 | Boronnitride | 10.95 | 3.09 | 24 | 24 | 80 | N |
| 66 | Hydrogencyanide | 13.01 | −2.59 | 14 | 14 | 98 | N |
| 67 | Phosphorusmononitride | 10.91 | −0.16 | 19 | 19 | 88 | N |

| | Name | IP | EA | $N_\omega$ | $N_\tau$ | $N_{bas}$ | fit set |
|-----|------|-----|-----|-----|-----|-----|-----|
| 68 | Hydrazene | 8.94 | −0.82 | 18 | 18 | 152 | G |
| 69 | Formaldehyde | 10.13 | −1.34 | 19 | 19 | 116 | G |
| 70 | Methanol | 10.32 | −1.05 | 18 | 18 | 152 | VG |
| 71 | Ethanol | 9.89 | −0.97 | 18 | 18 | 228 | VG |
| 72 | Acetaldehyde | 9.36 | −1.43 | 19 | 19 | 192 | VG |
| 73 | Ethoxyethane | 9.12 | −0.83 | 17 | 17 | 380 | VG |
| 74 | FormicAcid | 10.46 | −2.12 | 18 | 18 | 156 | G |
| 75 | Hydrogenperoxide | 10.69 | −2.38 | 18 | 18 | 116 | G |
| 76 | Water | 11.50 | −1.02 | 18 | 18 | 76 | N |
| 77 | Carbondioxide | 13.05 | −1.03 | 17 | 17 | 120 | N |
| 78 | Carbondisulfide | 9.42 | −0.17 | 19 | 19 | 136 | N |
| 79 | Carbonoxysulfide | 10.62 | −1.52 | 19 | 19 | 128 | N |
| 80 | Carbonoxyselenide | 9.88 | −1.25 | 24 | 24 | 151 | N |
| 81 | Carbonmonoxide | 13.23 | −1.16 | 18 | 18 | 80 | N |
| 82 | Ozon | 11.63 | 1.70 | 20 | 20 | 120 | N |
| 83 | Sulferdioxide | 11.61 | 0.58 | 20 | 20 | 128 | N |
| 84 | Berylliummonoxide | 9.16 | 1.85 | 20 | 20 | 80 | N |
| 85 | Magnesiummonoxide | 6.76 | 1.55 | 23 | 23 | 88 | N |
| 86 | Tuloene | 8.48 | −1.15 | 18 | 18 | 424 | G |
| 87 | Ethybenzene | 8.33 | −1.17 | 17 | 17 | 500 | G |
| 88 | Hexafluorobenzene | 9.26 | −0.46 | 17 | 17 | 480 | G |
| 89 | Phenol | 8.21 | −1.13 | 17 | 17 | 388 | G |
| 90 | Aniline | 7.49 | −1.30 | 18 | 18 | 406 | G |
| 91 | Pyridine | 8.82 | −0.77 | 16 | 16 | 330 | G |
| 92 | Guanine | 7.55 | −0.87 | 19 | 19 | 530 | VG |
| 93 | Adenine | 7.77 | −0.66 | 18 | 18 | 490 | G |
| 94 | Cytosine | 8.04 | −0.56 | 19 | 19 | 410 | G |
| 95 | Thymine | 8.51 | −0.36 | 19 | 19 | 468 | G |
| 96 | Uracil | 9.03 | −0.34 | 19 | 19 | 392 | VG |
| 97 | Urea | 8.99 | −0.64 | 18 | 18 | 232 | G |
| 98 | Silverdimer | 6.99 | 0.70 | 27 | 31 | 178 | N |
| 99 | Copperdimer | 7.13 | 0.64 | 24 | 24 | 126 | N |
| 100 | Coppercyanide | 9.58 | 1.08 | 24 | 24 | 143 | N |

Table 3: IPs and EAs on the aug-QZ6P level of theory (in eV) and technical parameters used in the calculations: Number of grid points, number of orbitals and fit set (N = *Normal*, G = *Good*, VG = *VeryGood*).

| | Name | IP | EA | $N_\omega$ | $N_\tau$ | $N_{bas}$ | fit set |
|---|---|---|---|---|---|---|---|
| 1 | Helium | 23.28 | −0.84 | 14 | 14 | 33 | N |
| 2 | Neon | 20.08 | −1.22 | 19 | 19 | 68 | N |
| 3 | Argon | 14.81 | −0.82 | 24 | 24 | 82 | N |
| 4 | Krypton | 13.41 | −0.75 | 27 | 31 | 120 | N |
| 5 | Xenon | 11.76 | −0.68 | 27 | 32 | 150 | N |
| 6 | Hydrogen | 15.72 | −0.90 | 14 | 14 | 66 | N |
| 7 | Lithiumdimer | 4.90 | 0.43 | 18 | 18 | 128 | N |
| 8 | Sodiumdimer | 4.79 | 0.50 | 24 | 24 | 146 | N |
| 9 | Sodiumtetramer | 4.19 | 0.92 | 24 | 24 | 292 | N |
| 10 | Sodiumhexamer | 4.24 | 0.95 | 24 | 24 | 438 | N |
| 11 | Dipotassium | 3.93 | 0.53 | 24 | 24 | 198 | N |
| 12 | Dirubidium | 3.76 | 0.59 | 27 | 28 | 274 | N |
| 13 | Nitrogen | 14.71 | −2.54 | 18 | 18 | 140 | N |
| 14 | Phosphorusdimer | 10.05 | 0.49 | 24 | 24 | 172 | N |
| 15 | Arsenicdimer | 9.28 | 0.74 | 24 | 24 | 238 | N |
| 16 | Fluorine | 14.78 | 0.15 | 21 | 21 | 140 | N |
| 17 | Chlorine | 10.93 | 0.56 | 24 | 24 | 174 | N |
| 18 | Bromine | 10.15 | 1.12 | 27 | 32 | 248 | N |
| 19 | Iodine | 9.02 | 1.36 | 27 | 31 | 274 | N |
| 20 | Methane | 13.84 | −0.89 | 17 | 17 | 196 | VG |
| 21 | Ethane | 12.28 | −0.88 | 18 | 18 | 326 | VG |
| 22 | Propane | 11.73 | −0.83 | 18 | 18 | 456 | VG |
| 23 | Buthane | 11.44 | −0.81 | 18 | 18 | 586 | VG |
| 24 | Ethylene | 10.17 | −2.03 | 18 | 18 | 260 | VG |
| 25 | Acetylene | 10.95 | −2.65 | 17 | 17 | 194 | VG |
| 26 | Tetracarbon | 10.63 | 2.40 | 22 | 22 | 256 | G |
| 27 | Cyclopropane | 10.49 | −0.88 | 18 | 18 | 390 | VG |
| 28 | Benzene | 8.92 | −1.14 | 19 | 19 | 582 | G |
| 29 | Cyclooctatetraene | 8.01 | −0.17 | 20 | 20 | 776 | G |
| 30 | Cyclopentadiene | 8.25 | −1.07 | 19 | 19 | 518 | G |
| 31 | Vynilfluoride | 10.09 | −2.09 | 20 | 20 | 297 | G |
| 32 | Vynilchloride | 9.62 | −1.49 | 24 | 24 | 314 | G |
| 33 | Vynilbromide | 8.87 | −1.39 | 27 | 32 | 351 | N |
| 34 | Vyniliodide | 8.86 | −0.95 | 27 | 32 | 372 | N |
| 35 | Carbontetrafluoride | 15.24 | −0.90 | 19 | 19 | 344 | G |
| 36 | Carbontetrachloride | 10.93 | −0.21 | 24 | 24 | 412 | G |
| 37 | Carbontetrabromide | 9.81 | 0.78 | 27 | 29 | 560 | N |
| 38 | Carbontetraiodide | 8.66 | 1.81 | 27 | 32 | 644 | N |
| 39 | Silane | 12.26 | −0.83 | 22 | 22 | 213 | VG |
| 40 | Germane | 11.95 | −0.77 | 24 | 24 | 251 | G |
| 41 | Disilane | 10.26 | −0.85 | 24 | 24 | 360 | G |
| 42 | Pentasilane | 8.88 | −0.26 | 24 | 24 | 801 | G |
| 43 | Lithiumhydride | 6.21 | 0.01 | 14 | 14 | 97 | N |
| 44 | Potassiumhydride | 4.85 | 0.11 | 24 | 24 | 132 | N |
| 45 | Borane | 12.78 | −0.39 | 17 | 17 | 163 | G |
| 46 | Diborane6 | 11.83 | −0.99 | 18 | 18 | 326 | G |
| 47 | Amonia | 10.12 | −0.85 | 19 | 19 | 169 | G |
| 48 | Hydrogenazide | 10.28 | −1.55 | 19 | 19 | 243 | G |
| 49 | Phosphine | 10.12 | −0.76 | 24 | 24 | 185 | G |

| | Name | IP | EA | $N_\omega$ | $N_\tau$ | $N_{bas}$ | fit set |
|---|---|---|---|---|---|---|---|
| 50 | Arsine | 10.06 | −0.73 | 24 | 24 | 218 | G |
| 51 | Hydrogensulfide | 9.90 | −0.85 | 24 | 24 | 148 | N |
| 52 | Hydrogenfluoride | 14.99 | −1.15 | 19 | 19 | 103 | N |
| 53 | Hydrogenchloride | 12.06 | −1.30 | 24 | 24 | 120 | N |
| 54 | Lithiumfluoride | 9.90 | −0.06 | 21 | 21 | 134 | N |
| 55 | Magnesiumfluoride | 12.27 | 0.15 | 20 | 20 | 213 | N |
| 56 | Titaniumfluoride | 13.78 | −0.48 | 27 | 27 | 384 | N |
| 57 | Aluminumtrifluoride | 14.16 | −0.10 | 22 | 22 | 291 | G |
| 58 | Fluoroborane | 10.36 | −1.38 | 21 | 21 | 134 | N |
| 59 | Sulfertetrafluoride | 12.02 | −0.55 | 24 | 24 | 362 | G |
| 60 | Potassiumbromide | 7.50 | 0.27 | 27 | 32 | 223 | N |
| 61 | Galliummonochloride | 9.50 | −0.13 | 24 | 24 | 206 | N |
| 62 | Sodiumchloride | 8.14 | 0.35 | 24 | 24 | 160 | N |
| 63 | Magnesiumchloride | 10.84 | 0.44 | 24 | 24 | 247 | N |
| 64 | Aluminumtriiodide | 9.24 | 0.64 | 27 | 31 | 516 | N |
| 65 | Boronnitride | 10.95 | 3.47 | 24 | 24 | 134 | N |
| 66 | Hydrogencyanide | 13.05 | −2.48 | 17 | 17 | 167 | N |
| 67 | Phosphorusmononitride | 10.99 | −0.04 | 24 | 24 | 156 | N |
| 68 | Hydrazene | 9.14 | −0.77 | 19 | 19 | 272 | G |
| 69 | Formaldehyde | 10.26 | −1.22 | 21 | 21 | 195 | G |
| 70 | Methanol | 10.41 | −0.95 | 19 | 19 | 261 | VG |
| 71 | Ethanol | 10.01 | −0.87 | 20 | 20 | 391 | VG |
| 72 | Acetaldehyde | 9.46 | −1.32 | 21 | 21 | 325 | VG |
| 73 | Ethoxyethane | 9.24 | −0.74 | 20 | 20 | 651 | VG |
| 74 | FormicAcid | 10.61 | −2.00 | 20 | 20 | 260 | G |
| 75 | Hydrogenperoxide | 10.80 | −2.25 | 19 | 19 | 196 | G |
| 76 | Water | 11.69 | −0.96 | 19 | 19 | 131 | N |
| 77 | Carbondioxide | 13.17 | −1.03 | 19 | 19 | 194 | N |
| 78 | Carbondisulfide | 9.58 | −0.06 | 24 | 24 | 228 | N |
| 79 | Carbonoxysulfide | 10.73 | −1.40 | 24 | 24 | 211 | N |
| 80 | Carbonoxyselenide | 10.08 | −1.13 | 24 | 24 | 248 | N |
| 81 | Carbonmonoxide | 13.39 | −1.04 | 20 | 20 | 129 | N |
| 82 | Ozon | 11.67 | 1.83 | 23 | 23 | 195 | N |
| 83 | Sulferdioxide | 11.71 | 0.69 | 24 | 24 | 212 | N |
| 84 | Berylliummonoxide | 9.10 | 1.90 | 23 | 23 | 129 | N |
| 85 | Magnesiummonoxide | 6.78 | 1.62 | 24 | 24 | 138 | N |
| 86 | Tuloene | 8.58 | −1.05 | 19 | 19 | 712 | G |
| 87 | Ethybenzene | 8.46 | −1.05 | 19 | 19 | 842 | G |
| 88 | Hexafluorobenzene | 9.44 | −0.29 | 21 | 21 | 804 | G |
| 89 | Phenol | 8.31 | −0.99 | 20 | 20 | 647 | G |
| 90 | Aniline | 7.58 | −1.14 | 20 | 20 | 685 | G |
| 91 | Pyridine | 8.96 | −0.64 | 19 | 19 | 555 | G |
| 92 | Guanine | 7.64 | −0.71 | 21 | 21 | 900 | VG |
| 93 | Adenine | 7.92 | −0.50 | 20 | 20 | 835 | G |
| 94 | Cytosine | 8.20 | −0.40 | 21 | 21 | 696 | G |
| 95 | Thymine | 8.65 | −0.20 | 20 | 20 | 788 | G |
| 96 | Uracil | 9.13 | −0.18 | 21 | 21 | 658 | VG |
| 97 | Urea | 9.10 | −0.58 | 20 | 20 | 401 | G |
| 98 | Silverdimer | 7.02 | 0.78 | 27 | 31 | 284 | N |
| 99 | Copperdimer | 7.56 | 0.78 | 27 | 32 | 210 | N |
| 100 | Coppercyanide | 9.85 | 1.24 | 27 | 29 | 239 | N |

Table 4: Ionization potentials and electron affinities for the subset of 250 molecules from the GW5000 database using TZ3P and QZ6P basis sets as well as complete basis set limit extrapolated values. All values are in eV.

| Name | IP | | | EA | | |
|------|------|------|-------|-------|------|-------|
|      | TZ3P | QZ6P | extra | TZ3P | QZ6P | extra |
| 16    | 7.20  | 7.39  | 7.63  | −0.20 | 0.05  | 0.36  |
| 212   | 8.01  | 8.14  | 8.30  | 0.34  | 0.59  | 0.90  |
| 389   | 8.29  | 8.46  | 8.66  | 0.33  | 0.55  | 0.84  |
| 584   | 10.30 | 10.43 | 10.60 | 1.44  | 1.65  | 1.92  |
| 964   | 8.16  | 8.22  | 8.30  | 0.27  | 0.50  | 0.78  |
| 1145  | 7.30  | 7.48  | 7.69  | 0.51  | 0.73  | 1.01  |
| 1304  | 8.58  | 8.70  | 8.86  | 1.27  | 1.44  | 1.65  |
| 1415  | 7.83  | 8.00  | 8.22  | −0.30 | −0.02 | 0.33  |
| 1627  | 7.50  | 7.70  | 7.95  | 1.34  | 1.56  | 1.84  |
| 1761  | 8.05  | 8.24  | 8.48  | 0.12  | 0.35  | 0.65  |
| 1942  | 7.77  | 7.92  | 8.10  | −0.29 | −0.05 | 0.24  |
| 2142  | 8.07  | 8.26  | 8.49  | 0.57  | 0.82  | 1.14  |
| 2403  | 7.14  | 7.32  | 7.54  | 0.65  | 0.86  | 1.12  |
| 2686  | 7.71  | 7.87  | 8.08  | 1.16  | 1.38  | 1.66  |
| 2869  | 8.58  | 8.71  | 8.86  | 0.80  | 1.00  | 1.25  |
| 3133  | 10.76 | 10.90 | 11.06 | 1.51  | 1.74  | 2.03  |
| 3387  | 7.51  | 7.68  | 7.88  | 0.50  | 0.73  | 1.01  |
| 3793  | 8.71  | 8.88  | 9.08  | −0.06 | 0.20  | 0.53  |
| 4002  | 7.93  | 8.06  | 8.23  | 2.01  | 2.20  | 2.45  |
| 4257  | 8.23  | 8.32  | 8.45  | −0.10 | 0.17  | 0.54  |
| 4465  | 7.87  | 8.09  | 8.36  | 0.32  | 0.56  | 0.86  |
| 4727  | 7.73  | 7.92  | 8.16  | −0.18 | 0.07  | 0.37  |
| 4986  | 8.52  | 8.64  | 8.80  | −0.17 | 0.13  | 0.51  |
| 5179  | 7.16  | 7.34  | 7.58  | 0.83  | 1.07  | 1.37  |
| 5330  | 8.28  | 8.47  | 8.70  | 0.35  | 0.57  | 0.84  |
| 5760  | 10.49 | 10.61 | 10.75 | 0.13  | 0.41  | 0.74  |
| 5948  | 6.73  | 6.91  | 7.14  | −0.10 | 0.17  | 0.49  |
| 6247  | 7.74  | 7.89  | 8.08  | −0.08 | 0.17  | 0.47  |
| 6527  | 8.48  | 8.65  | 8.87  | 1.00  | 1.23  | 1.53  |
| 6838  | 8.71  | 8.87  | 9.08  | −0.13 | 0.10  | 0.39  |
| 7071  | 7.37  | 7.53  | 7.72  | 0.17  | 0.40  | 0.68  |
| 7348  | 8.07  | 8.33  | 8.66  | −0.18 | 0.05  | 0.33  |
| 7474  | 8.14  | 8.32  | 8.54  | 0.45  | 0.68  | 0.96  |
| 7729  | 8.66  | 8.80  | 8.96  | 0.64  | 0.86  | 1.14  |
| 7902  | 8.04  | 8.26  | 8.52  | 0.07  | 0.35  | 0.70  |
| 8115  | 7.90  | 8.08  | 8.29  | 0.53  | 0.78  | 1.10  |
| 8314  | 8.88  | 9.06  | 9.27  | −0.26 | 0.01  | 0.34  |
| 8509  | 8.93  | 9.10  | 9.32  | 0.19  | 0.42  | 0.71  |
| 8740  | 7.93  | 8.12  | 8.36  | −0.02 | 0.26  | 0.62  |
| 9040  | 6.85  | 6.97  | 7.12  | −0.22 | −0.04 | 0.19  |
| 9202  | 8.11  | 8.28  | 8.48  | 0.91  | 1.10  | 1.33  |
| 9538  | 8.36  | 8.55  | 8.79  | 1.27  | 1.48  | 1.73  |
| 9844  | 8.35  | 8.48  | 8.64  | 0.08  | 0.36  | 0.71  |
| 10214 | 10.05 | 10.20 | 10.38 | 2.06  | 2.25  | 2.48  |
| 10450 | 7.64  | 7.79  | 7.97  | 0.05  | 0.30  | 0.60  |
| 10698 | 9.21  | 9.41  | 9.65  | 0.54  | 0.78  | 1.08  |
| 10978 | 7.17  | 7.35  | 7.57  | −0.10 | 0.15  | 0.46  |
| 11151 | 7.89  | 8.05  | 8.25  | 0.55  | 0.77  | 1.05  |
| 11403 | 11.14 | 11.26 | 11.41 | −0.40 | −0.17 | 0.12  |
| 11661 | 8.33  | 8.43  | 8.56  | −0.18 | 0.07  | 0.37  |
| 12004 | 7.39  | 7.58  | 7.81  | 0.15  | 0.39  | 0.70  |

Continued on next page

8

| Name | IP TZ3P | QZ6P | extra | EA TZ3P | QZ6P | extra |
|------|------|------|-------|------|------|-------|
| 12143 | 8.25 | 8.43 | 8.65 | 0.52 | 0.74 | 1.01 |
| 12405 | 7.89 | 8.03 | 8.20 | 0.13 | 0.37 | 0.66 |
| 12569 | 9.32 | 9.46 | 9.64 | 0.20 | 0.43 | 0.72 |
| 12919 | 7.51 | 7.72 | 7.97 | 0.18 | 0.42 | 0.73 |
| 13151 | 7.82 | 7.98 | 8.17 | −0.02 | 0.21 | 0.49 |
| 13321 | 7.88 | 8.03 | 8.21 | 1.04 | 1.28 | 1.58 |
| 13505 | 7.49 | 7.66 | 7.88 | 0.17 | 0.41 | 0.70 |
| 13702 | 6.72 | 6.91 | 7.14 | −0.35 | −0.06 | 0.30 |
| 13712 | 7.71 | 7.86 | 8.04 | 0.82 | 1.02 | 1.26 |
| 13722 | 7.85 | 8.05 | 8.30 | 0.29 | 0.52 | 0.80 |
| 13736 | 7.03 | 7.18 | 7.36 | 0.00 | 0.19 | 0.43 |
| 13760 | 7.17 | 7.33 | 7.54 | 0.19 | 0.46 | 0.80 |
| 14098 | 6.89 | 7.09 | 7.33 | 0.34 | 0.58 | 0.87 |
| 14226 | 7.59 | 7.78 | 8.01 | −0.06 | 0.18 | 0.48 |
| 14670 | 7.98 | 8.17 | 8.41 | 0.48 | 0.73 | 1.05 |
| 14979 | 8.66 | 8.85 | 9.08 | −0.24 | 0.07 | 0.47 |
| 15273 | 7.35 | 7.50 | 7.68 | 0.01 | 0.25 | 0.54 |
| 15429 | 6.40 | 6.58 | 6.80 | 0.58 | 0.84 | 1.15 |
| 15634 | 8.29 | 8.46 | 8.67 | −0.27 | −0.03 | 0.27 |
| 15938 | 7.85 | 8.00 | 8.18 | 0.08 | 0.36 | 0.71 |
| 16245 | 7.45 | 7.68 | 7.96 | −0.12 | 0.14 | 0.47 |
| 16444 | 8.20 | 8.39 | 8.63 | 0.62 | 0.84 | 1.11 |
| 16704 | 6.21 | 6.37 | 6.58 | 0.93 | 1.16 | 1.45 |
| 16849 | 7.78 | 7.97 | 8.19 | 0.08 | 0.32 | 0.62 |
| 16982 | 7.66 | 7.81 | 8.02 | 0.11 | 0.31 | 0.57 |
| 17264 | 7.14 | 7.30 | 7.50 | 0.78 | 1.04 | 1.37 |
| 17502 | 8.37 | 8.53 | 8.73 | 0.40 | 0.64 | 0.95 |
| 17807 | 7.46 | 7.53 | 7.62 | −0.02 | 0.16 | 0.39 |
| 18111 | 7.01 | 7.14 | 7.31 | 0.15 | 0.39 | 0.69 |
| 18255 | 8.65 | 8.79 | 8.97 | 0.96 | 1.20 | 1.49 |
| 18460 | 6.97 | 7.19 | 7.46 | 0.75 | 0.98 | 1.28 |
| 18611 | 8.54 | 8.69 | 8.88 | −0.28 | −0.01 | 0.33 |
| 18825 | 7.18 | 7.35 | 7.58 | 0.47 | 0.71 | 1.02 |
| 19062 | 7.57 | 7.75 | 7.98 | −0.03 | 0.23 | 0.57 |
| 19347 | 8.36 | 8.45 | 8.57 | 1.64 | 1.88 | 2.18 |
| 19664 | 8.57 | 8.76 | 9.00 | 0.69 | 0.98 | 1.34 |
| 19910 | 9.48 | 9.61 | 9.77 | 1.03 | 1.27 | 1.56 |
| 20065 | 7.90 | 8.06 | 8.26 | 0.58 | 0.80 | 1.08 |
| 20311 | 8.63 | 8.78 | 8.97 | −0.17 | 0.08 | 0.38 |
| 20649 | 7.67 | 7.81 | 7.99 | −0.23 | 0.01 | 0.29 |
| 20821 | 8.15 | 8.33 | 8.56 | 0.82 | 1.06 | 1.36 |
| 21105 | 8.06 | 8.24 | 8.46 | 1.96 | 2.16 | 2.41 |
| 21210 | 7.92 | 8.07 | 8.26 | 0.13 | 0.37 | 0.67 |
| 21361 | 8.24 | 8.38 | 8.56 | 1.20 | 1.42 | 1.69 |
| 21611 | 6.53 | 6.69 | 6.90 | 0.50 | 0.73 | 1.02 |
| 21895 | 9.28 | 9.42 | 9.59 | 0.03 | 0.25 | 0.52 |
| 22078 | 7.20 | 7.34 | 7.53 | 1.24 | 1.47 | 1.76 |
| 22407 | 8.66 | 8.81 | 9.00 | 1.19 | 1.42 | 1.70 |
| 22699 | 8.15 | 8.29 | 8.45 | 1.19 | 1.40 | 1.66 |
| 22875 | 7.39 | 7.58 | 7.83 | −0.17 | 0.07 | 0.38 |
| 23028 | 8.66 | 8.79 | 8.96 | 0.39 | 0.60 | 0.86 |
| 23303 | 7.34 | 7.48 | 7.66 | 1.31 | 1.51 | 1.76 |
| 23652 | 7.86 | 8.02 | 8.23 | 0.17 | 0.42 | 0.72 |
| 23853 | 7.41 | 7.59 | 7.81 | 0.10 | 0.34 | 0.64 |
| 24031 | 6.23 | 6.40 | 6.62 | 0.11 | 0.37 | 0.70 |
| 24201 | 6.68 | 6.87 | 7.10 | −0.19 | 0.06 | 0.36 |

| Name | IP | | | EA | | |
|------|------|------|-------|------|------|-------|
| | TZ3P | QZ6P | extra | TZ3P | QZ6P | extra |
| 24419 | 8.52 | 8.69 | 8.92 | −0.17 | 0.08 | 0.40 |
| 24722 | 7.11 | 7.26 | 7.46 | 0.10 | 0.33 | 0.61 |
| 24951 | 7.36 | 7.52 | 7.73 | 0.22 | 0.47 | 0.78 |
| 25240 | 9.45 | 9.64 | 9.88 | −0.14 | 0.10 | 0.41 |
| 25412 | 7.74 | 7.83 | 7.94 | 0.47 | 0.71 | 1.00 |
| 25789 | 8.60 | 8.77 | 8.98 | 0.08 | 0.32 | 0.62 |
| 25995 | 7.21 | 7.39 | 7.61 | 0.01 | 0.25 | 0.55 |
| 26246 | 7.83 | 7.99 | 8.18 | 0.69 | 0.90 | 1.17 |
| 26458 | 7.58 | 7.75 | 7.97 | −0.04 | 0.21 | 0.51 |
| 26685 | 7.82 | 7.99 | 8.20 | 0.41 | 0.65 | 0.94 |
| 26821 | 7.93 | 8.07 | 8.24 | 0.12 | 0.36 | 0.66 |
| 27000 | 7.46 | 7.65 | 7.90 | −0.02 | 0.22 | 0.53 |
| 27374 | 7.80 | 7.95 | 8.13 | −0.08 | 0.16 | 0.46 |
| 27595 | 9.37 | 9.50 | 9.66 | 1.05 | 1.24 | 1.49 |
| 27801 | 7.81 | 8.00 | 8.23 | −0.08 | 0.20 | 0.53 |
| 28006 | 7.50 | 7.67 | 7.89 | 0.33 | 0.57 | 0.88 |
| 28162 | 7.23 | 7.33 | 7.45 | −0.01 | 0.15 | 0.35 |
| 28450 | 7.33 | 7.42 | 7.53 | 0.73 | 0.90 | 1.11 |
| 28674 | 8.08 | 8.23 | 8.41 | −0.08 | 0.18 | 0.51 |
| 28988 | 7.56 | 7.77 | 8.03 | 0.15 | 0.43 | 0.78 |
| 29288 | 9.34 | 9.43 | 9.55 | −0.09 | 0.20 | 0.57 |
| 29484 | 7.88 | 8.07 | 8.30 | 0.42 | 0.67 | 0.97 |
| 29738 | 7.84 | 8.00 | 8.22 | −0.17 | 0.08 | 0.40 |
| 30014 | 6.81 | 6.98 | 7.20 | 0.91 | 1.16 | 1.47 |
| 30240 | 7.10 | 7.30 | 7.55 | −0.19 | 0.06 | 0.37 |
| 30510 | 6.64 | 6.86 | 7.13 | 0.10 | 0.37 | 0.71 |
| 30647 | 7.72 | 7.86 | 8.04 | 0.01 | 0.26 | 0.58 |
| 30833 | 7.59 | 7.78 | 8.01 | 0.75 | 0.97 | 1.23 |
| 31114 | 7.56 | 7.75 | 7.99 | 0.25 | 0.49 | 0.78 |
| 31332 | 7.79 | 7.94 | 8.14 | 0.24 | 0.49 | 0.81 |
| 31529 | 7.69 | 7.87 | 8.08 | −0.22 | −0.03 | 0.19 |
| 31853 | 7.67 | 7.74 | 7.83 | 0.69 | 0.85 | 1.06 |
| 32294 | 6.55 | 6.76 | 7.03 | 1.36 | 1.61 | 1.91 |
| 32571 | 8.97 | 9.10 | 9.27 | −0.09 | 0.18 | 0.52 |
| 32947 | 7.09 | 7.20 | 7.33 | 0.62 | 0.82 | 1.06 |
| 33146 | 9.04 | 9.18 | 9.35 | −0.08 | 0.16 | 0.46 |
| 33372 | 7.48 | 7.61 | 7.75 | −0.19 | 0.02 | 0.29 |
| 33531 | 7.43 | 7.57 | 7.74 | 0.29 | 0.52 | 0.80 |
| 33692 | 8.40 | 8.56 | 8.77 | 0.42 | 0.67 | 0.99 |
| 34005 | 7.40 | 7.58 | 7.81 | 0.08 | 0.36 | 0.70 |
| 34307 | 7.42 | 7.61 | 7.84 | 0.97 | 1.23 | 1.56 |
| 34564 | 6.52 | 6.72 | 6.96 | 0.42 | 0.68 | 1.00 |
| 34913 | 7.08 | 7.24 | 7.45 | 0.28 | 0.52 | 0.84 |
| 35225 | 7.99 | 8.12 | 8.29 | −0.23 | 0.06 | 0.43 |
| 35442 | 6.91 | 7.07 | 7.27 | 0.53 | 0.78 | 1.11 |
| 35790 | 7.20 | 7.37 | 7.58 | 0.16 | 0.46 | 0.82 |
| 36205 | 8.35 | 8.50 | 8.70 | 0.72 | 0.96 | 1.26 |
| 36515 | 7.76 | 7.98 | 8.26 | 0.35 | 0.60 | 0.92 |
| 36735 | 7.15 | 7.33 | 7.55 | 1.10 | 1.34 | 1.64 |
| 37128 | 8.40 | 8.55 | 8.74 | 1.94 | 2.11 | 2.32 |
| 37381 | 8.17 | 8.31 | 8.48 | −0.26 | −0.02 | 0.28 |
| 37765 | 9.23 | 9.36 | 9.51 | −0.06 | 0.16 | 0.43 |
| 38018 | 7.08 | 7.27 | 7.51 | 0.14 | 0.40 | 0.72 |
| 38315 | 8.49 | 8.66 | 8.87 | 0.41 | 0.64 | 0.93 |
| 38639 | 6.41 | 6.58 | 6.79 | 0.07 | 0.32 | 0.63 |
| 38920 | 6.89 | 7.06 | 7.26 | 0.69 | 0.94 | 1.24 |

| Name | IP | | | EA | | |
| --- | --- | --- | --- | --- | --- | --- |
| | TZ3P | QZ6P | extra | TZ3P | QZ6P | extra |
| 39175 | 6.59 | 6.79 | 7.05 | 0.02 | 0.29 | 0.62 |
| 39418 | 7.94 | 8.12 | 8.34 | 0.66 | 0.91 | 1.21 |
| 39685 | 7.75 | 7.92 | 8.14 | 0.82 | 1.02 | 1.27 |
| 39917 | 7.80 | 7.93 | 8.09 | −0.19 | 0.06 | 0.36 |
| 40143 | 7.87 | 8.08 | 8.34 | 0.66 | 0.91 | 1.22 |
| 40494 | 7.88 | 8.04 | 8.26 | 0.39 | 0.61 | 0.90 |
| 40764 | 8.44 | 8.56 | 8.70 | 1.73 | 1.96 | 2.25 |
| 40978 | 8.71 | 8.89 | 9.12 | −0.38 | −0.15 | 0.14 |
| 41377 | 8.24 | 8.38 | 8.57 | 0.13 | 0.40 | 0.74 |
| 41571 | 6.44 | 6.62 | 6.85 | 0.55 | 0.83 | 1.17 |
| 41897 | 7.69 | 7.90 | 8.15 | 0.65 | 0.77 | 0.93 |
| 42090 | 7.54 | 7.74 | 7.99 | 0.15 | 0.42 | 0.75 |
| 42424 | 7.82 | 8.03 | 8.29 | −0.23 | 0.05 | 0.40 |
| 42754 | 7.05 | 7.24 | 7.47 | 0.45 | 0.70 | 1.01 |
| 42908 | 7.20 | 7.40 | 7.64 | 0.25 | 0.49 | 0.80 |
| 43090 | 6.59 | 6.70 | 6.85 | 0.83 | 1.01 | 1.24 |
| 43385 | 7.30 | 7.47 | 7.68 | 0.49 | 0.73 | 1.04 |
| 43634 | 7.76 | 7.96 | 8.21 | 0.20 | 0.45 | 0.74 |
| 43905 | 7.45 | 7.66 | 7.93 | 0.07 | 0.31 | 0.62 |
| 44205 | 7.17 | 7.23 | 7.31 | 0.28 | 0.40 | 0.56 |
| 44586 | 6.15 | 6.31 | 6.51 | 0.57 | 0.78 | 1.04 |
| 44870 | 8.15 | 8.33 | 8.57 | −0.03 | 0.22 | 0.53 |
| 45218 | 8.04 | 8.22 | 8.45 | 0.23 | 0.42 | 0.66 |
| 45485 | 7.91 | 8.10 | 8.35 | 1.13 | 1.37 | 1.67 |
| 45666 | 8.63 | 8.76 | 8.92 | −0.12 | 0.14 | 0.47 |
| 45995 | 7.67 | 7.88 | 8.15 | 0.26 | 0.52 | 0.84 |
| 46362 | 8.57 | 8.72 | 8.91 | −0.20 | 0.03 | 0.32 |
| 46610 | 8.42 | 8.59 | 8.81 | 0.01 | 0.30 | 0.67 |
| 46821 | 6.97 | 7.14 | 7.35 | 0.71 | 0.94 | 1.22 |
| 46991 | 8.69 | 8.81 | 8.97 | 0.47 | 0.72 | 1.03 |
| 47200 | 7.04 | 7.15 | 7.27 | 0.32 | 0.57 | 0.87 |
| 47575 | 7.68 | 7.87 | 8.12 | 0.54 | 0.80 | 1.13 |
| 47797 | 7.10 | 7.32 | 7.59 | 0.46 | 0.73 | 1.06 |
| 47960 | 5.93 | 6.17 | 6.47 | 1.15 | 1.39 | 1.70 |
| 48162 | 8.03 | 8.18 | 8.36 | 0.81 | 1.06 | 1.37 |
| 48399 | 7.68 | 7.86 | 8.08 | 0.91 | 1.13 | 1.40 |
| 48653 | 7.49 | 7.65 | 7.84 | 1.29 | 1.50 | 1.75 |
| 48947 | 6.89 | 7.09 | 7.35 | −0.10 | 0.14 | 0.44 |
| 49106 | 7.64 | 7.81 | 8.01 | 0.66 | 0.91 | 1.22 |
| 49471 | 6.01 | 6.19 | 6.42 | 1.46 | 1.70 | 2.00 |
| 49946 | 7.66 | 7.83 | 8.03 | 0.26 | 0.50 | 0.80 |
| 50224 | 8.39 | 8.55 | 8.75 | 0.35 | 0.61 | 0.95 |
| 50401 | 7.72 | 7.91 | 8.14 | −0.14 | 0.10 | 0.41 |
| 50771 | 7.76 | 7.91 | 8.09 | −0.08 | 0.16 | 0.45 |
| 51045 | 7.49 | 7.70 | 7.97 | −0.07 | 0.14 | 0.39 |
| 51317 | 6.98 | 7.18 | 7.43 | 0.40 | 0.66 | 0.98 |
| 51639 | 7.67 | 7.83 | 8.02 | 1.04 | 1.25 | 1.51 |
| 51981 | 8.30 | 8.45 | 8.64 | 0.21 | 0.46 | 0.78 |
| 52259 | 8.23 | 8.41 | 8.62 | 0.48 | 0.76 | 1.10 |
| 52590 | 6.50 | 6.69 | 6.92 | 0.69 | 0.82 | 0.98 |
| 52978 | 8.17 | 8.33 | 8.52 | −0.40 | −0.15 | 0.15 |
| 53229 | 8.52 | 8.65 | 8.81 | 0.16 | 0.47 | 0.86 |
| 53566 | 9.70 | 9.77 | 9.86 | −0.41 | −0.15 | 0.16 |
| 53842 | 8.08 | 8.22 | 8.40 | 0.96 | 1.17 | 1.43 |
| 54009 | 7.61 | 7.78 | 8.00 | −0.27 | −0.02 | 0.30 |
| 54233 | 7.60 | 7.77 | 7.97 | −0.24 | 0.00 | 0.31 |

| Name | IP TZ3P | QZ6P | extra | EA TZ3P | QZ6P | extra |
|---|---|---|---|---|---|---|
| 54412 | 8.48 | 8.62 | 8.80 | 0.89 | 1.10 | 1.36 |
| 54680 | 7.16 | 7.41 | 7.72 | −0.04 | 0.21 | 0.52 |
| 54908 | 6.95 | 7.15 | 7.39 | 1.60 | 1.80 | 2.06 |
| 55110 | 8.31 | 8.49 | 8.73 | −0.10 | 0.15 | 0.45 |
| 55259 | 7.57 | 7.73 | 7.93 | 0.66 | 0.87 | 1.13 |
| 55516 | 8.98 | 9.12 | 9.31 | 0.19 | 0.42 | 0.72 |
| 55803 | 8.93 | 9.09 | 9.29 | 0.64 | 0.87 | 1.14 |
| 56050 | 8.72 | 8.86 | 9.04 | −0.18 | 0.07 | 0.39 |
| 56219 | 8.67 | 8.88 | 9.14 | 0.58 | 0.81 | 1.10 |
| 56406 | 7.22 | 7.39 | 7.60 | 0.46 | 0.70 | 0.98 |
| 56584 | 9.88 | 9.99 | 10.14 | 1.05 | 1.26 | 1.52 |
| 56782 | 8.59 | 8.72 | 8.87 | 0.13 | 0.37 | 0.65 |
| 57147 | 7.10 | 7.28 | 7.49 | 0.37 | 0.59 | 0.87 |
| 57383 | 9.26 | 9.37 | 9.50 | 1.02 | 1.22 | 1.46 |
| 57610 | 7.95 | 8.11 | 8.31 | −0.27 | 0.01 | 0.35 |
| 57896 | 7.59 | 7.68 | 7.80 | 0.54 | 0.76 | 1.04 |
| 58206 | 6.85 | 7.02 | 7.24 | 0.60 | 0.83 | 1.12 |
| 58443 | 8.82 | 8.96 | 9.13 | −0.24 | 0.01 | 0.34 |
| 58653 | 9.14 | 9.30 | 9.49 | 0.09 | 0.32 | 0.61 |
| 58846 | 8.03 | 8.15 | 8.30 | −0.01 | 0.23 | 0.53 |
| 59124 | 7.06 | 7.23 | 7.43 | 1.14 | 1.33 | 1.57 |
| 59304 | 9.51 | 9.60 | 9.71 | 1.41 | 1.58 | 1.78 |
| 59631 | 7.32 | 7.47 | 7.65 | 0.58 | 0.80 | 1.06 |
| 59849 | 8.58 | 8.76 | 8.99 | −0.04 | 0.21 | 0.51 |
| 60181 | 7.80 | 7.95 | 8.13 | 0.74 | 0.97 | 1.25 |
| 60360 | 8.46 | 8.69 | 8.95 | 0.44 | 0.66 | 0.92 |
| 60545 | 7.77 | 7.95 | 8.18 | 0.13 | 0.36 | 0.64 |
| 60749 | 7.67 | 7.82 | 8.01 | 0.78 | 1.01 | 1.31 |
| 60961 | 7.86 | 7.97 | 8.10 | 0.97 | 1.19 | 1.46 |
| 61133 | 7.17 | 7.35 | 7.56 | 0.13 | 0.35 | 0.63 |
| 61346 | 7.74 | 7.90 | 8.11 | −0.05 | 0.21 | 0.54 |

# Graphical TOC Entry

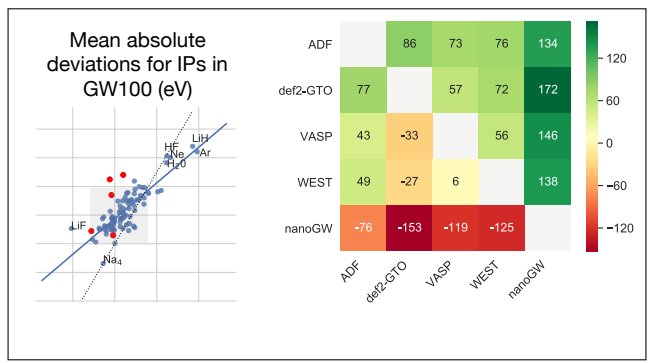



\end{document}